\pdfoutput=1

\documentclass[11pt]{article}

\usepackage[final]{acl}

\usepackage{times}
\usepackage{latexsym}

\usepackage[T1]{fontenc}

\usepackage[utf8]{inputenc}

\usepackage{microtype}

\usepackage{inconsolata}

\usepackage{graphicx, booktabs, amsmath}

\usepackage{hyperref}
\usepackage{booktabs}
\usepackage{adjustbox}
\usepackage{wrapfig}
\usepackage{amssymb}
\usepackage{color,xcolor,colortbl}
\usepackage{enumitem}
\usepackage{soul}
\usepackage{amsmath}
\usepackage{url}
\usepackage{algorithm}
\usepackage[noend]{algpseudocode}
\usepackage{booktabs}
\usepackage{bm,bbm}
\usepackage{mathtools}
\usepackage{array}
\usepackage{multirow}
\usepackage{subcaption}
\usepackage{pbox}
\usepackage{pifont}
\usepackage{arydshln}
\usepackage{dblfloatfix}
\usepackage{relsize}
\usepackage{setspace}
\usepackage[scaled=.8]{beramono}
\usepackage{tcolorbox}
\usepackage{listings}

\newtcolorbox{promptbox}[2][]{
  floatplacement={#2},
  colframe=dark,colback=light!30!white,
  fonttitle=\small\ttfamily,
  fontupper=\small\ttfamily,
  title=#2,
  boxrule=0.5mm, 
  halign=flush left,
}

\definecolor{dark}{HTML}{012169}
\definecolor{light}{HTML}{efede1}

\usepackage{authblk}

\title{\textsc{Harbor}: Exploring Persona Dynamics in Multi-Agent Competition}

\author{\textbf{Kenan Jiang}, \textbf{Li Xiong}, \textbf{Fei Liu}}
\affil[]{Emory University}
\affil[]{\texttt{\{kenan.jiang, lxiong, fei.liu\}@emory.edu}}

\begin{document}
\maketitle
\begin{abstract}
We investigate factors contributing to LLM agents' success in competitive multi-agent environments, using auctions as a testbed where agents bid to maximize profit. The agents are equipped with bidding domain knowledge, distinct personas that reflect item preferences, and a memory of auction history. Our work extends the classic auction scenario by creating a realistic environment where multiple agents bid on houses, weighing aspects such as size, location, and budget to secure the most desirable homes at the lowest prices. Particularly, we investigate three key questions: (a) How does a persona influence an agent's behavior in a competitive setting? (b) Can an agent effectively profile its competitors' behavior during auctions? (c) How can persona profiling be leveraged to create an advantage using strategies such as theory of mind? Through a series of experiments, we analyze the behaviors of LLM agents and shed light on new findings.  Our testbed, called \textsc{Harbor}, offers a valuable platform for deepening the understanding of multi-agent workflows in competitive environments. 

\end{abstract}

\section{Introduction}

When everyone has an LLM agent by their side, it is crucial these agents help users make decisions that align with their personal preferences~\cite{frisch2024llmagentsinteractionmeasuring,eigner2024determinantsllmassisteddecisionmaking}. For example, they may act as proxies for job candidates competing for limited positions or assist home buyers in strategizing their purchases~\cite{an2024largelanguagemodelsdiscriminate}. Candidates have unique skills and job preferences, while home buyers have distinct preferences and budgets~\cite{samuel2024personagymevaluatingpersonaagents}. Without thoroughly understanding the persona dynamics shaping LLMs' behavior in competitive environments, designing optimal strategies can be challenging.

Existing research largely focuses on enhancing LLM agents' core capabilities, such as reasoning, planning, tool use, grounding, and multi-modal perception~\cite{qin2023toolllmfacilitatinglargelanguage,valmeekam2023planningabilitieslargelanguage,bohnet2024exploringbenchmarkingplanningcapabilities,li2024laspsurveyingstateoftheartlarge,li2025embodiedagentinterfacebenchmarking}. They also seek to improve agents' interactions with external environments, such as navigating the web or physical world, using APIs, querying databases, or retrieving documents~\cite{zhou2024webarenarealisticwebenvironment,xie2024osworldbenchmarkingmultimodalagents,xu2024theagentcompanybenchmarkingllmagents, agashe2025agent}. Some studies have explored research from a multi-agent perspective~\cite{Li_2023,guo2024largelanguagemodelbased,zhang2024llmmastermindsurveystrategic}. We are particularly interested in the direction where the agents must seamlessly integrate personalized preferences and anticipate others' behaviors to compete effectively. 

We present \textsc{Harbor}, a new testbed for studying persona dynamics in competitive environments. \textsc{Harbor} simulates real house bidding, where buyers' preferences, budgets, and competitors' choices significantly influence purchasing outcomes. Unlike prior studies of games such as the Prisoner's Dilemma or poker~\cite{yim2024evaluating,wang2024unleashingemergentcognitivesynergy,hua2024gametheoreticllmagentworkflow}, in which agents lack individual preferences and outcomes are based on Nash equilibria, our research emphasizes persona dynamics among multiple agents. We profile various buyer types, from first-time homebuyers to flippers and downsizers, using real data from \href{https://www.redfin.com}{Redfin.com}. Further, our work extends human-human negotiation conversations~\cite{he-etal-2018-decoupling,yang-etal-2021-improving,dutt-etal-2021-resper,lin-etal-2024-decision} into a multi-agent setting to assess the strengths and weaknesses of agents equipped with personas.

\begin{figure*}
    \centering
    \includegraphics[width=5.8in]{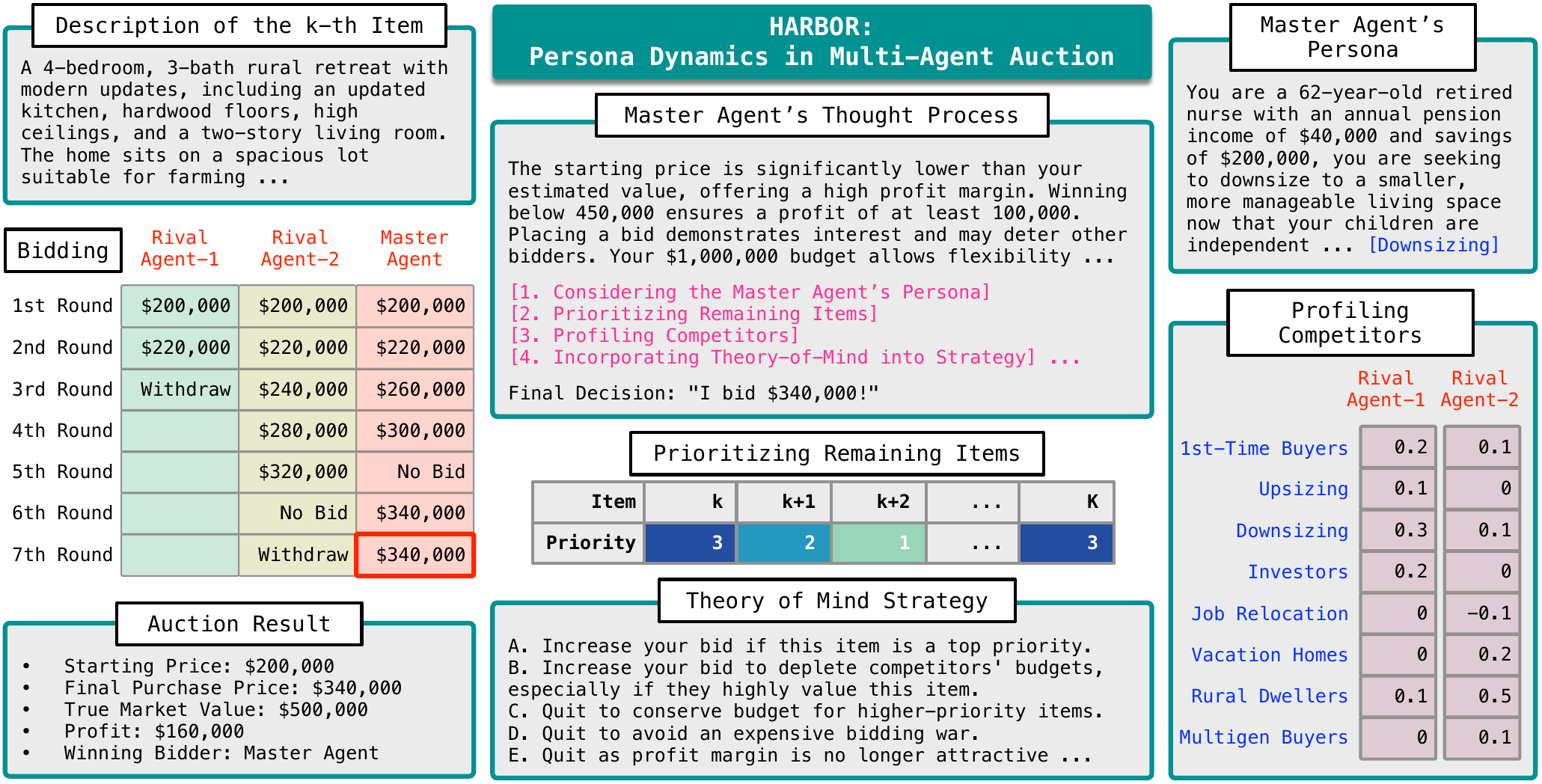}
    \caption{Our \textsc{Harbor} platform explores persona dynamics in multi-agent auctions. The Master Agent learns to prioritize items, profile competitors based on their bidding behavior, and make bidding decisions using theory of mind strategies. Through a series of experiments, we analyze agent behavior and uncover new insights.}
    \label{fig:example}
\vspace{-0.1in}
\end{figure*}

Our platform enables the analysis of multi-agent decision-making in competitive settings. When bidding for multiple items, agents must plan their actions, decide which items to prioritize, manage budgets, and sometimes give up lower-priority items to secure higher-value ones~\cite{chen2023put}. Agents must also analyze competitors' behavior and adapt their tactics accordingly. This paper does not focus on training multi-agent RL systems~\cite{anonymous2025singleagent}. Instead, we explore how injecting personas into LLM agents shapes their behavior in auctions. We examine how aggressively an agent bids (e.g., the number of attempts and amount raised) and how personas influence bidding outcomes, measured by profitability and success in securing persona-aligned items. These results have important implications for competitive scenarios, such as companies bidding for contracts, advertisers competing for ad space, or individuals negotiating deals. Our contributions in this paper are summarized as follows:

\begin{itemize}[topsep=3pt,itemsep=-1pt,leftmargin=*]

\item \textsc{Harbor} enhances auction dynamics by incorporating personas, allowing us to study LLM agents beyond traditional game theory. We explore three key questions: (a) How does a persona influence an agent's behavior? (b) Can an agent effectively profile its competitors' behavior during auctions? (c) How can persona profiling help in developing strategies such as the theory of mind?

\item We introduce a new approach to evaluate agent performance, combining persona profiling (via KL divergence) with competitiveness metrics: profitability and TrueSkill~\cite{herbrich2007trueskill}. Profitability measures an agent's gains relative to all possible profit margins, while TrueSkill factors in both wins and opponent competitiveness. These metrics help us better understand agent behavior and uncover valuable insights.\footnote{Our source code will be shared publicly upon acceptance.}

\end{itemize}

\section{Related Work}
\label{sec:related}

\paragraph{Multi-Agent Orchestration}\; Multi-agent systems are now drawing increasing attention. Notable examples include OpenAI's Agents SDK~\cite{openai_agents_sdk}, Microsoft's AutoGen~\cite{microsoft_autogen}, and Google's AI Co-Scientist~\cite{gottweis2025ai}. This momentum is also fueled by the Model Context Protocol (MCP; \citealt{anthropic_mcp}), which lets agents interact with external data and tools, and Agent2Agent (A2A; \citealt{google_a2a_2025}), which enables agents to collaborate and delegate tasks. So far, many studies have explored the collaborative potential of multi-agent systems~\cite{gu2024agent,li2024autokagglemultiagentframeworkautonomous,zhang2024chainagentslargelanguage,tran2025multiagentcollaborationmechanismssurvey,liao2025marftmultiagentreinforcementfinetuning}.

We are especially interested in what capabilities make an agent stand out in competitive multi-agent settings~\cite{chen2024llmarenaassessingcapabilitieslarge,zhao2024competeai,wang2024battleagentbenchbenchmarkevaluatingcooperation,wu-etal-2024-deciphering,geng2025realmbenchrealworldplanningbenchmark,zhu2025multiagentbenchevaluatingcollaborationcompetition}. In this line of work, \citet{zhang2024llmmastermindsurveystrategic} explored social simulations and board games, finding that even advanced models struggle to fully utilize their reasoning potential. \citet{huang2024far} introduced an evaluation metric to assess LLMs' gaming abilities in multi-agent settings through game-theoretic experiments. \citet{zhu2025multiagentbenchevaluatingcollaborationcompetition} proposed a benchmark to explore agent coordination protocols and strategies, such as group discussion and cognitive planning.

Our work extends AucArena~\cite{chen2023put} by introducing persona dynamics into a bidding framework where multiple agents compete for maximum profit in an auction. Our framework aims for a deeper investigation of an agent's ability to profile the behavior of others. Furthermore, we explore how theory of mind influences agent performance in a competitive multi-agent setting.

\vspace{-0.05in}
\paragraph{Agent Personas}\;
Personas function as identities assigned to LLM agents, enhancing their ability to generate personalized or specialized outputs~\cite{xu2024character, chen2024from, sun2024building, li2025helloagainllmpoweredpersonalized}. Prior research has extensively examined the influence of personas across various roles~\cite{hu-collier-2024-quantifying, samuel2024personagymevaluatingpersonaagents, kim2024persona, dong-etal-2024-llm}. Leveraging their role-playing nature, some studies have applied personas to social simulations by assigning diverse identities to entire agent populations~\cite{lee2024language, tseng-etal-2024-two, hu2024quantifying}. \citet{yang2024} find that persona injection can introduce shortcut learning, causing LLM agents to deviate from rational objectives.  
Building on this, we incorporate persona into a multi-agent auction to examine its effects in competitive settings.

\vspace{-0.05in}
\paragraph{Theory of Mind (ToM) Strategies}\;\;
Theory of Mind, the ability to understand and infer one's own and others' mental states, is fundamental to human social interaction and a crucial capability for LLMs to achieve human-like reasoning~\cite{leslie2004core, sap-etal-2022-neural, van-duijn-etal-2023-theory, li2023camel, light2023avalonbench, cross2024hypothetical, chan-etal-2024-negotiationtom}. An accurate ToM in others' intentions and actions provides significant advantages~\cite{street2024llm, amirizaniani2024llms}. Past studies have applied ToM to simulate social behaviors. \citet{de2017negotiating} explored its role in negotiation tasks, while \citet{wang-etal-2024-sotopia} proposed an interactive learning environment to train LLMs in social interactions. In contrast, our work investigates whether ToM can enhance agent performance in competing multi-agent environments with auction as the testbed. 

\section{The \textsc{Harbor} Framework}
\label{sec:framework}

We now introduce \textsc{Harbor} (\textbf{H}ousing \textbf{A}uction for \textbf{R}easoning, \textbf{B}idding, and \textbf{O}pponent \textbf{R}ecognition), our framework for studying how persona-driven agents reason about components and apply theory of mind in multi-agent settings. It provides a realistic and flexible environment for agent interaction, with customizable agents and auction items to reflect individual preferences. We also include measurable metrics for evaluating agent performance.

\subsection{Auction Environment Setup}
\label{sec:setup}

Multiple agents $N$ compete for a series of items $H$ through an open bidding process. Agents can observe each other's actions in real time. While these agents are profit-driven, they may also be assigned personas. These personas shape bidding behaviors by creating preferences for specific items while still prioritizing overall profit maximization.

Each agent has access to the complete list of items $H$, including their publicly announced starting prices $V^0_{h\in H}$ and item descriptions. However, the true values of items $V^*_{h\in H}$ remain hidden from the agents. Instead, each agent estimates an item's worth based on an overhead percentage applied to the true value. For instance, if an item has a known starting price of \$200 but an undisclosed true value of \$500, and the agent's overhead estimation is 10\%, it will perceive the item's value as \$550. In the bidding process, agents must begin at the stated starting price and can either place a higher bid than the current leading offer or withdraw from that round. The bidding continues until only one agent remains, at which point it secures ownership of the item, and the final bid amount is deducted from the winner's budget. The profit and the maximum possible profit for item $h$ is computed as:

\begin{equation}
    \text{Profit} = V^*_h - \bar{V}_h
\end{equation}
\begin{equation}
    \text{max Profit} = V^*_h - V^0_h
\end{equation}
where \( \bar{V}_h \) represents the winning bid paid only by the agent who wins the item. The auction concludes once all items have undergone the bidding process.

\begin{figure*}
    \centering
    \includegraphics[width=6.2in]{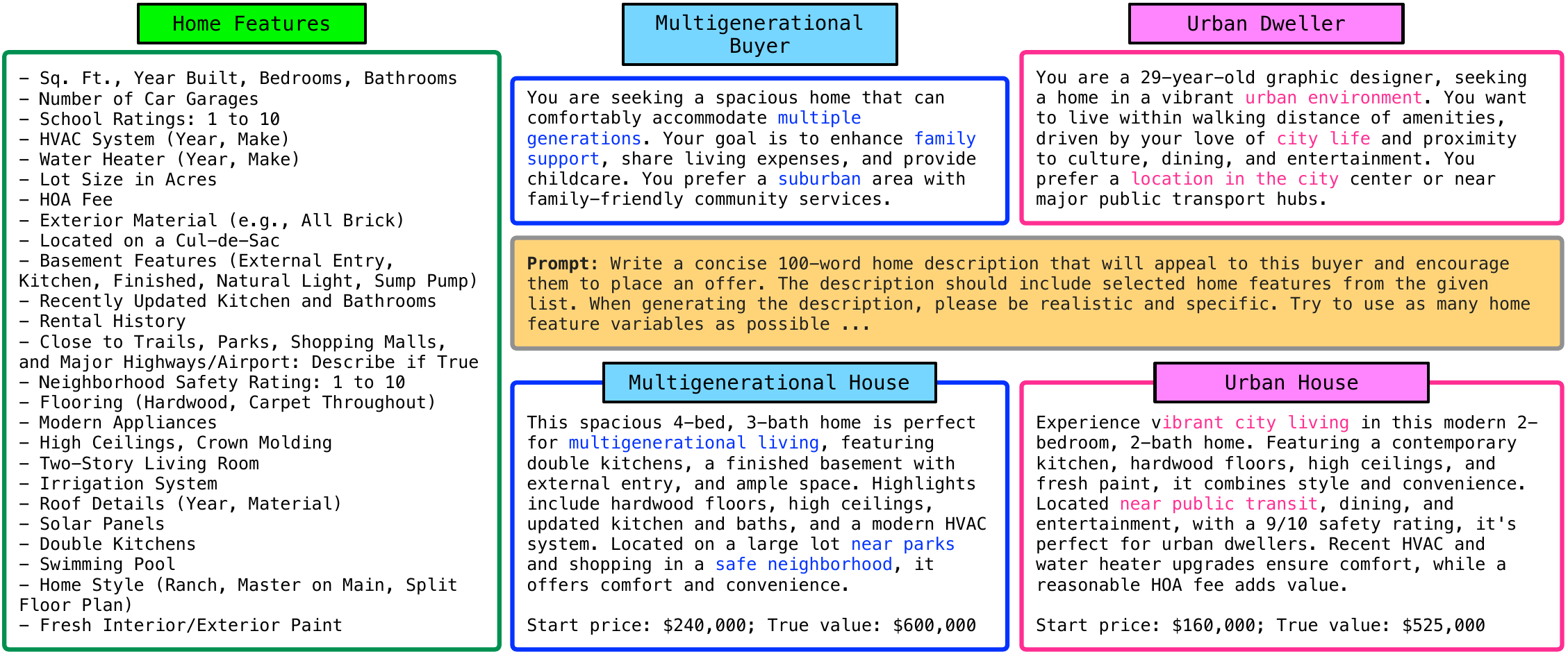} 
    \vspace{-0.05in}
    \caption{Example Bidders (top) and their Houses (bottom). We curate a diverse set of houses tailored to different personas. A list of home features and personas is gathered from \href{https://www.redfin.com}{Redfin.com} to enhance data realism.}
    \label{fig: persona_house}
\vspace{-0.15in}
\end{figure*}

\subsection{Priority-Based Planning}
\label{sec:planning}

During an auction, an agent constructs and continuously updates a \textbf{priority list} \( L \), a dynamic rating system that assigns a score \( l_h \in \{1,2,3\} \) to each item \( h \), where 3 represents the highest priority and 1 the lowest. This evolving priority list guides the agent's bidding decisions throughout the auction. The agent places more bids on high-priority items and may withdraw on low-priority ones.

Before bidding, the agent initializes \( L_0 \) based on its initial budget \( B_0 \), persona \( \pi \), available items \( H_0 \), and the objective of maximizing the profit ratio, \( \max R \). Each item \( h \in H_0 \) is assigned an initial priority score in a single prompt, generating the initial priority list \( L_0 \) with length \( |H_0| \):
\[
L_0 = LLM(\{l_h\}_{h\in H_0}|B_0, \pi, H_0, \max R)
\]
This ensures that the agent starts the auction with a well-defined strategic priority structure aligned with its objectives.  

After completing the bidding process for an item \( h_t \), the agent updates its priority list \( L_{t-1} \) to reflect its new planning. This update is determined by the agent's remaining budget \( B_t \), the set of available items \( H_t \). It also incorporates the status of all agents, denoted as \( S_t \), which has their acquired items and profits. Additionally, the update considers \( P_t \), a collection of vectors containing the agent's estimate of its competitors' personas.  
The priority list update process can be formalized as:  
\[
L_{t} = LLM(\{l_h\}_{h\in H_t}|B_t, \pi, H_t, S_t, P_t, L_{t-1})
\]
This iterative process enables agents to continuously refine their strategies in real time, adjusting their decisions to better align with their objectives. 

\subsection{Agent Persona Modeling}
\label{sec:persona}

Agents in competitive settings might not act with pure objectivity; instead, they adapt to human preferences while pursuing their assigned goals~\cite{tseng-etal-2024-two}. For instance, real estate agents that act as surrogates for buyers may prefer certain homes and aim to secure them at the lowest prices. Our \textsc{Harbor} is designed to simulate such realistic auctions by incorporating diverse personas that reflect real-world motivations. 

We ask a domain expert to create a range of home buyer profiles, such as first-time buyers, downsizing homeowners, and urban dwellers, using information from \url{Redfin.com}. An example of these profiles is shown in Figure~\ref{fig: persona_house}. Each LLM agent can adopt a single persona or a blend of two personas. Following established practices in persona and role-playing research \cite{chuang-etal-2024-beyond, park2023generative}, we implement persona injection by adding bidder-specific persona prompts to the system message \cite{xie2024can, li-etal-2025-hello, weissburg-etal-2025-llms}. 

For each agent persona, we generate persona-aligned home descriptions by prompting \texttt{GPT-4o} to ``\emph{write a concise 100-word home description that will appeal to this buyer and encourage them to place an offer. Use as many of the home features as possible...}'' These features include square footage, year built, number of bedrooms and bathrooms, lot size, and more. The LLM also provides both a starting price and the home’s estimated true value. Figure~\ref{fig: persona_house} shows two example bidders and their matched homes. \textsc{Harbor} thus enables diverse auction simulations by sampling from different personas and their corresponding homes. 

Our profiling module maintains a \( k \)-dimensional vector \( P^c \) for each competitor \( c \), where \( k \) represents the number of personas. Each value \( p^c \in P^c \) ranges from \([-1,1]\), indicating the weight of a persona, with higher values indicating stronger alignment. Before the auction begins, the agent initializes each competitor's profiling vector \( P^c \) as a zero vector. At the end of bidding round \( t \), the agent updates \( P^c_{t-1} \) based on the current item \( h \) and its bidding history \( T_h \), which records all agents' bidding actions for \( h \). If a competitor bids heavily on $h$, the profiling module increases weights for personas likely drawn to $h$; otherwise, it decreases them, potentially assigning negative weights. The profiling update process can be formalized as: 
\[
P^c_t = LLM(\{p^c_{t, i}\}_{ \forall i \in [1,k] } | P^c_{t-1}, h, T_h)
\]
Aggregating all vectors $P^c_t$ forms the complete profiling knowledge $P_t$ at round \( t \). This profiling knowledge is a key input in adjusting the agent's priority list (\S~\ref{sec:planning}). The profiling prompt and an example output are provided in Appendix~\ref{sec:profile_prompt}.

\subsection{Bidding Strategy Design (ToM)}
\label{sec:strategy}
A strategy guides the agent on how to act based on the current state of the auction. For example, a strategy might be ``\emph{increase the bid to drain competitors' budgets, especially if a competitor highly values this item}'' or simply, ``\emph{quit to conserve budget for higher-priority items}.'' Theory-of-mind refers to the ability infer both one's own and others' mental states. Our agent uses ToM to infer the personas of competing agents and incorporates that knowledge into its strategy design. The strategy module takes the current auction state as input and selects an appropriate action $a_t^*$. The auction state includes: (a) the status of all agents, $S_{t-1}$, including their acquired items and profits; (b) profile knowledge, $P_{t-1}$, which consists of vectors estimating the personas of competitors; (c) a priority list, $L_{t-1}$, showing the remaining items and their assigned priorities; and (d) the bidding context $C_t$ such as item details and the current bid price. Our \textbf{ExpertAct} strategy chooses from a predefined set containing six expert-specified rules (Appendix~\ref{sec:strategy_prompt}). In contrast, \textbf{RLAct} uses reinforcement learning to generate a reasoning trace and recommend whether to keep bidding (Bd) or withdraw (Wd).
\[
a_t^* = LLM( a_t \in \mbox{\small\{Bd, Wd\}} |P_{t-1}, L_{t-1}, S_{t-1}, C_t)
\]
For \textbf{RLAct}, we fine-tune \texttt{Llama-3-8B-Instruct} to obtain a reasoning system for bidding decisions. This setup lets us directly compare RL-based strategies (learned from data) with expert-crafted strategies (rich in domain knowledge). Due to resource constraints, we fine-tune the model using 1k training examples via online DPO~\cite{rafailov2024directpreferenceoptimizationlanguage}. At each bidding turn, the system gathers the full auction state, e.g., the system message, and the bidder's profile and transaction history. Then, we sample multiple candidate responses from the Llama model; each analyzes the situation and proposes a bidding action. We use \texttt{GPT-4.1-mini} as an online judge. It scores each candidate on a 1-5 scale (1 = worst, 5 = best). We focus on response pairs with large score differences and fine-tune the model to favor the higher-quality responses. At inference time, RLAct generates a reasoning trace based on the auction state and makes a decision to either bid or withdraw. More fine-tuning details are in Appendix~\ref{sec:dpo}.

\begin{figure*}[htbp]
    \centering
    \includegraphics[width=0.88\linewidth]{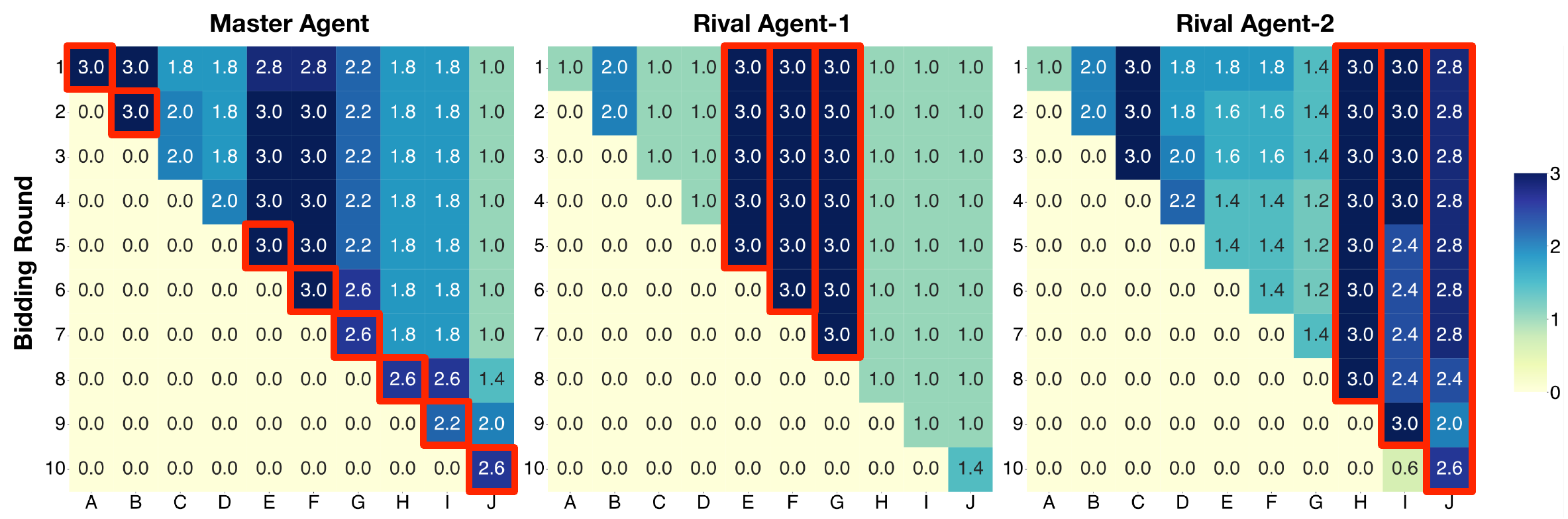} 
    \vspace{-0.05in}
    \caption{Heatmaps of Priority of Master Agent (left), Rival Agents 1 (middle) and 2 (right). Agent 1's persona matches to Item \{E, F, G\}. Agent 2's persona matches to Item \{H, I, J\}. High-priority scores highlighted in red.}
    \label{fig: experiment_1_big} 
    \vspace{-0.1in}
\end{figure*}

\subsection{Performance Evaluation}
\label{sec:eval}

We assess an agent's performance from multiple angles, combining persona profiling (via KL divergence) with competitiveness metrics. Profitability measures an agent's gains relative to all possible profit margins, while TrueSkill~\cite{herbrich2007trueskill} factors in both wins and opponent competitiveness. A top-performing agent not only understands their rivals' preferences, but also leverages that insight to develop strategies.

\vspace{-0.05in}
\paragraph{Agent Competitiveness}
The \textbf{Profit Ratio} measures an agent's earnings relative to the maximum possible profit, which is achieved by winning all items at their starting prices. This metric evaluates an agent's ability to maximize financial gains while accounting for price and item variations across auctions. \textbf{TrueSkill Score} ranks agents based on profitability relative to competitors; an agent with lower absolute profit can still achieve a high TrueSkill score if its earnings exceed those of competitors. Lastly, the \textbf{Matched-Item Acquisition Rate} quantifies the proportion of persona-aligned items successfully acquired, assessing the agent's effectiveness in adhering to its persona-driven objectives. Together, these metrics provide a robust framework for quantifying both persona alignment and strategic performance.

At the conclusion of the auction, let \( H_b \) denote the set of items won by agent \( b \), and let \( H_b^\pi \) represent the subset of items that align perfectly with agent \( b \)'s persona \( \pi \). Define \( I_{b,h} \in \{0,1\} \) as a binary indicator, where \( I_{b,h} = 1 \) if agent \( b \) wins item \( h \), and \( I_{b,h} = 0 \) otherwise. The performance metrics for agent \( b \) are defined as follows:

\begin{equation}
R_b = \frac{\sum_{h \in H} I_{b, h}\left( V^*_h - \bar{V}_h\right)}{\sum_{h \in H}\left( V^*_h - V^0_h\right)}
\end{equation}

\begin{equation}
A_b = \frac{\sum_{h \in H^\pi_b} I_{b, h}}{|H_b^\pi|}
\end{equation}
where \textbf{Profit Ratio} \( R_b \) and \textbf{Matched-Item Acquisition Rate} \( A_b \) quantifies the profit-driven and persona-driven objectives of agent \( b \) repectively.

In an auction setting with agents $b_1$, $b_2$, and $b_3$, the agent's \textbf{TrueSkill Score} $S_{b_i}$ is directly proportional to the ranking of the agents' profitability.
\[
S_{b_i} \propto \operatorname{rank}(R_i : i \in \{b_1, b_2, b_3\})
\]

\paragraph{Persona Profiling Accuracy}
This provides a rigorous measure of how well the profiling module captures a competitor's persona.
At the end of the auction, profiling performance is measured by comparing inferred profile vector \( P^c \) against ground truth persona vector \( G^c \). For \textbf{single-persona agents}, \( G^c \) is a one-hot vector with \( G^c_i = 1 \) in the competitor $c$'s true persona dimension. Similarly, for \textbf{mix-of-two persona agents}, \( G^c \) has two active dimensions, each weighted at \( 0.5 \). To ensure a positive probability distribution, we shift \( P^c \) by adding \( |\min(P^c)| \) and normalize it. A smoothing factor \( \epsilon = 10^{-12} \) is applied to both \( G^c \) and \( P^c \). Profiling accuracy is then evaluated using KL divergence:  
\[
D_{KL}(G^c \parallel P^c) = \sum G^c \log \frac{G^c}{P^c}.
\]

\section{Experiments}
\label{sec:experiments}

We conduct a series of experiments to address three key questions: (a) How does a persona influence an agent's behavior in competitive settings? (b) Can the agent effectively profile its competitors' behavior? (c) How can persona profiling be leveraged to create an advantage using strategies such as theory of mind? These experiments aim to improve our understanding of persona dynamics in multi-agent competitive scenarios.
Our `Master' agent undergoes various modifications and competes against a number of `Rival' agents. Each agent is assigned a persona that shapes its prioritization of items. To enhance robustness, each experiment runs five times, and we report the average results. All agents use the same foundational LLM; further experiment details are in Appendix~\ref{sec:exp_details}.

\subsection{Impact of Persona on Agent Behavior}
\label{sec:4.1}
This experiment demonstrates how an assigned persona influences an agent's behavior in auctions. We assign a unique persona to each of Rival 1 and Rival 2. Among ten items, each rival has three items matching their assigned persona. The Master Agent operates without a persona for comparison. 

Figure~\ref{fig: experiment_1_big} shows heatmaps of priority scores for the items across bidding rounds. The Master Agent, lacking a persona, determines item priority based on auction order, giving the highest priority to the first remaining items. Rival Agents 1 and 2 consistently assign the highest priority scores (3) to their matched items across all bidding rounds. 

\begin{figure}[t]
\centering
\includegraphics[width=0.85\linewidth]{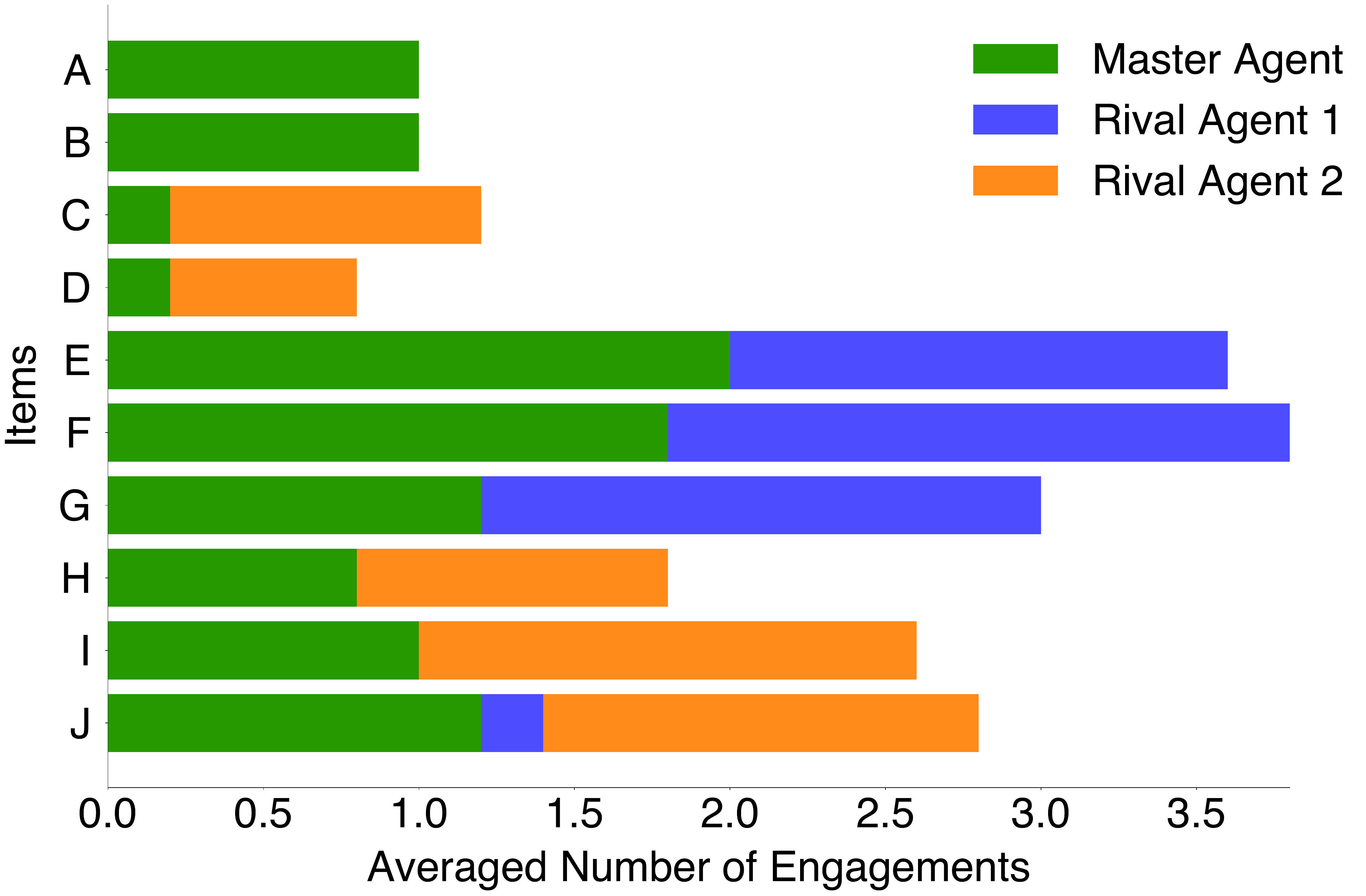} 
\vspace{-0.05in}
\caption{Engagements of all Items. Agent 1's persona matches to Item \{E, F, G\}. Agent 2's persona matches to Item \{H, I, J\}.}
\label{fig: engagement}
\vspace{-0.15in}
\end{figure}

Figure~\ref{fig: engagement} further supports this pattern by showing the average number of engagements, measured by how often each agent raises bids for an item. The results illustrate that Rivals 1 and 2 consistently favor items aligned with their personas. They assign high-priority scores to these items and repeatedly increase their bids. Our findings indicate that persona injection influences agents' decisions and diverts them from rational, profit-driven bidding.

\begin{figure*}
    \centering
    \includegraphics[width=0.33\linewidth]{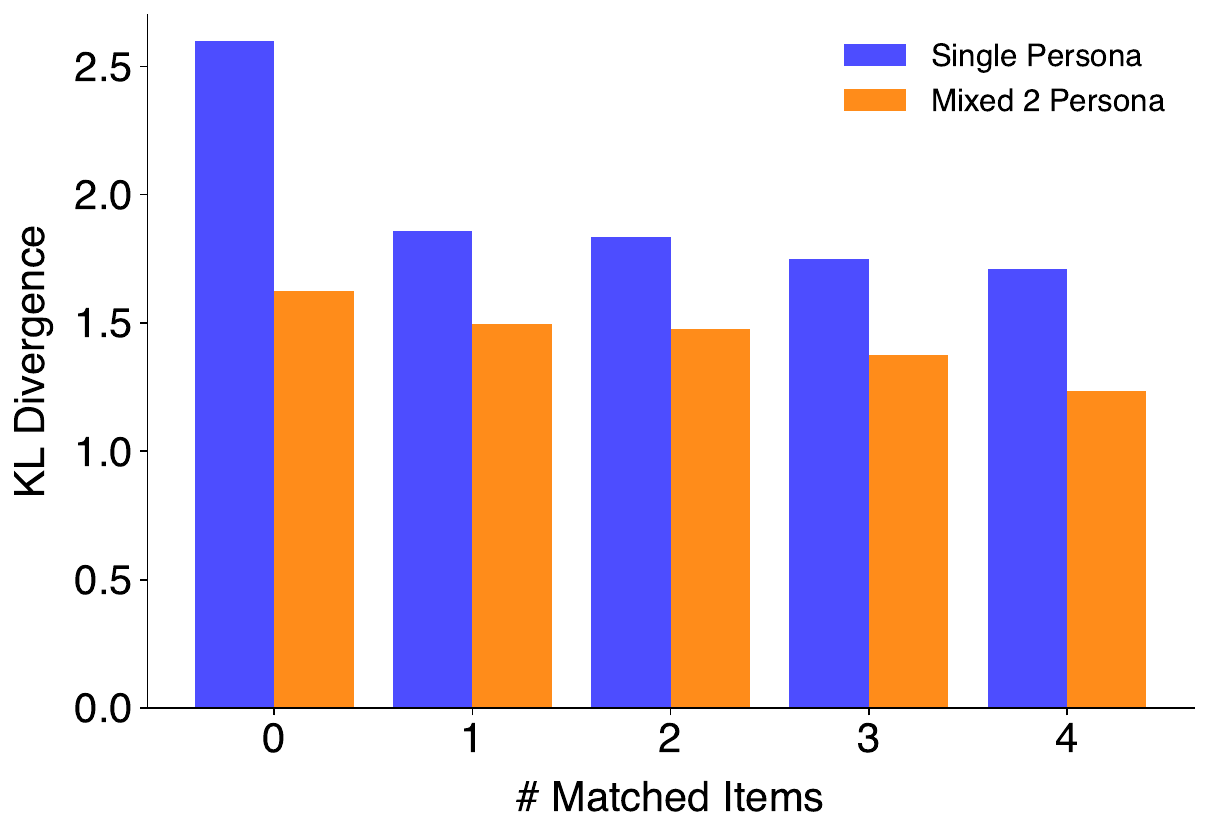} 
    \includegraphics[width=0.33\linewidth]{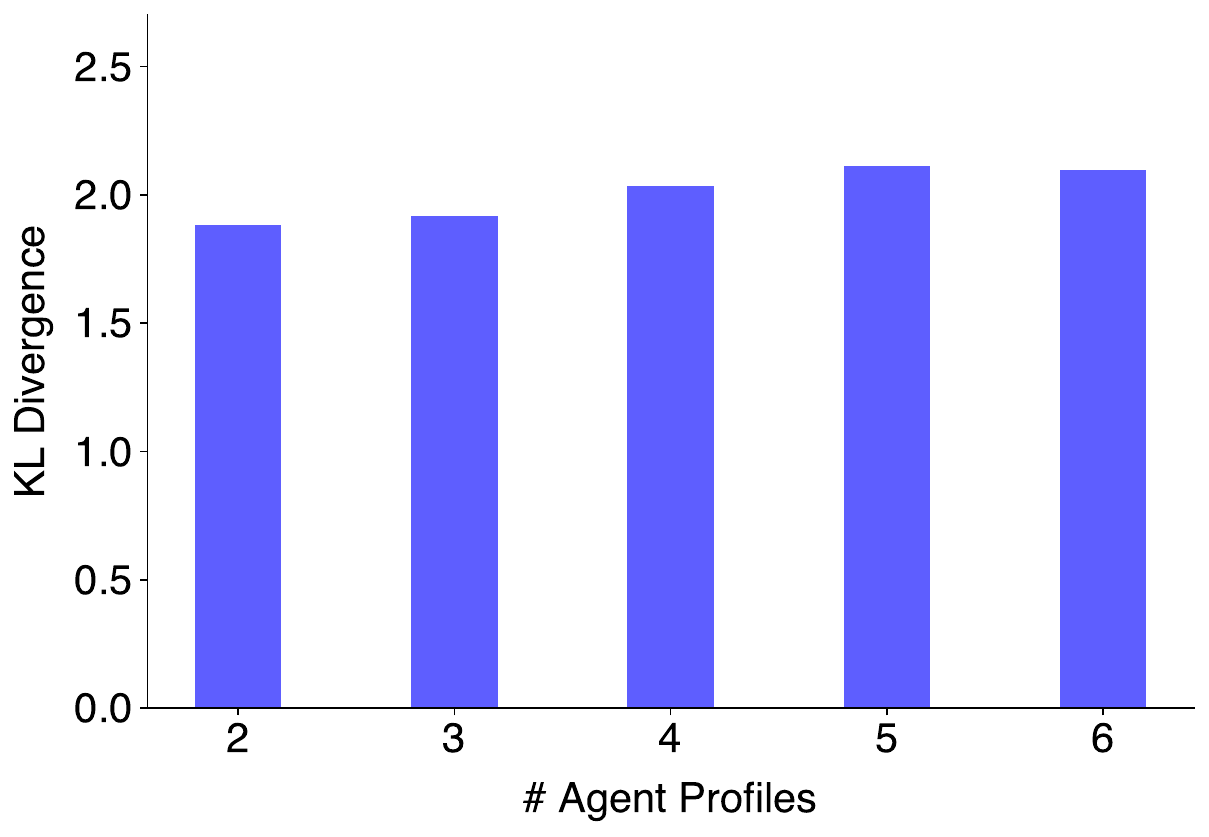} 
    \includegraphics[width=0.32\linewidth]{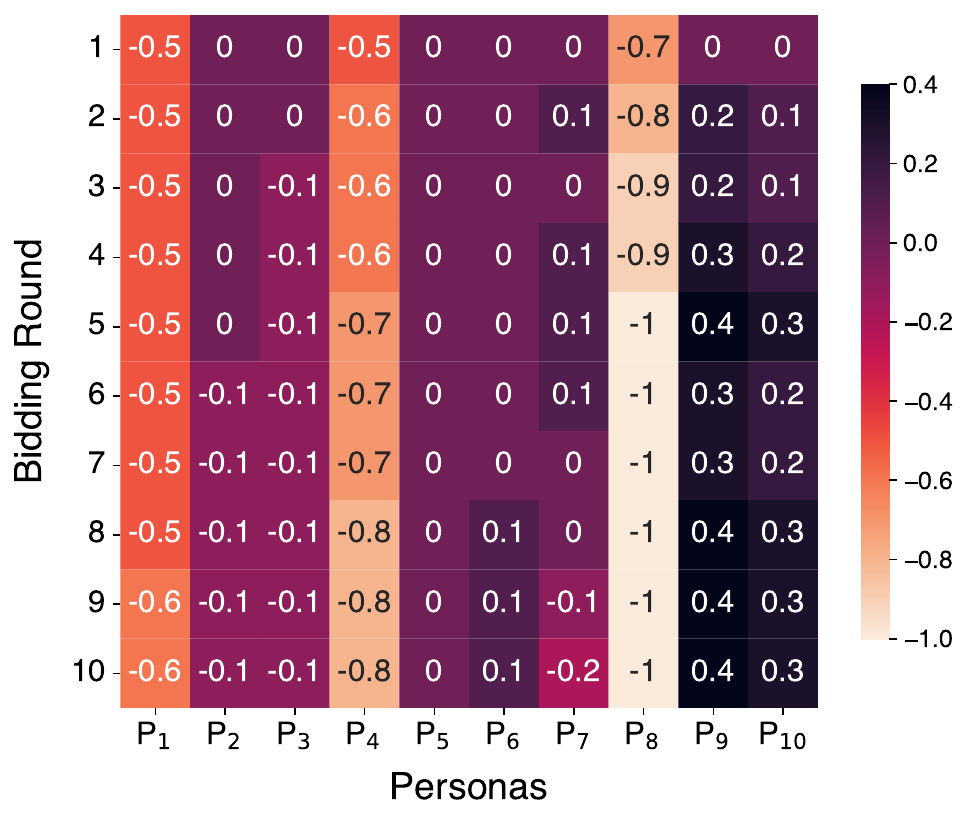} 
    \vspace{-0.05in}
    \caption{
    (\textsc{Left}) The KL divergence scores illustrate the Master Agent's ability to infer competitors' personas; (\textsc{Middle}) As more agents join (\# Agent Profiles), the Master struggles to infer personas accurately, as reflected by increasing KL divergence scores; (\textsc{Right}) Master Agent's profiling of a Rival evolves after each bidding round.
    }
    \label{fig:profile_persona}
\vspace{-0.1in}
\end{figure*}

\subsection{Impact of Persona on Profitability}
\label{sec:4.2}
Does assigning a persona to an agent affect its profitability? How does competition impact earnings when multiple agents share the same persona? All agents start with the same budget. Persona agents have two aligned items. The Master Agent is tested under four conditions: (a) \textbf{Master w/o Persona}: the Master Agent has no persona; (b) \textbf{Master w/ Persona}: the Master has a persona different from both Rival Agents; (c) \textbf{Some Competition}: the Master shares a persona with one rival (Rival 2), creating direct competition for persona-aligned items; (d) \textbf{More Competition}: all agents share the same persona, intensifying competition. To ensure results are persona-independent, we rotate agent personas in a round-robin manner and report the averaged results.

Results in Table~\ref{tab:experiment_1.1} suggest that persona injection affects the agent's core profit-driven objective. The Master Agent prioritizes items that align with its persona, making it less competitive for higher-profit items and leading to lower overall profits. Without a persona, the Master Agent achieves the highest profit rate of 34.56\%. Increased competition among agents further reduces overall profit rates. This effect is illustrated in the average profit rate decline from 29.23\% to 21.95\%. These results support the common intuition that entering a profit-driven auction with emotional or preferential biases can undermine a bidder's financial success.

\subsection{Profiling Rival Personas}
\label{sec:4.3}
To evaluate profiling accuracy, we use KL divergence to measure the difference between the inferred and ground truth persona distributions. Ground truth for single-persona agents is a one-hot vector, while mixed-persona agents distribute weights equally across two dimensions. Lower KL divergence implies more accurate persona inferences by the Master, while higher values suggest greater deviation.

We explore whether personas become more evident as more items align with an agent's persona. If an agent consistently bids on preferred items, their persona should be fully revealed. To test this, we analyze two Rival agents with their number of matched houses $M$ ranging from 0 to 4. For example, at $M$ = 4, each Rival Agent has four matched houses; at $M$ = 0, none matches their personas. The Master Agent, without a persona, aims to infer the personas of both Rivals.

\begin{table}[t]
\centering
\setlength{\tabcolsep}{3pt}
\renewcommand{\arraystretch}{1.1}
\begin{small}
\begin{tabular}{lcccc}
& \multicolumn{4}{c}{\textbf{Profit Rate ($R$)}} \\
\cmidrule(lr){2-5} 
\textbf{Setting} & \textbf{Master} & \textbf{Rival 1} & \textbf{Rival 2} & \textbf{Average}\\ 
\toprule
{Master w/o Persona} &      \textbf{34.56} & 24.59 & 24.08 &   27.74    \\ 
{Master w/ Persona} &     \textbf{31.75}  &   32.35    &   23.60      &     29.23  \\ 
{Some Competition} &   \textbf{26.81}    &  29.72      &  22.88    &    26.47     \\ 
{More Competition} &     \textbf{25.22}   &   21.66       &   18.97      & 21.95 \\ 
\bottomrule
\end{tabular}
\end{small}
\vspace{-0.05in}
\caption{Profit rates (\%) of the agents under four conditions. The Master Agent prioritizes persona-aligned items, reducing competitiveness for higher-profit items and lowering overall profits.}
\label{tab:experiment_1.1}
\vspace{-0.15in}
\end{table}

\begin{table*}[h]
\centering
\setlength{\tabcolsep}{4.3pt}
\renewcommand{\arraystretch}{1.1} 
\begin{small}
\begin{tabular}{lccccccccc}
\toprule
& \multicolumn{3}{c}{\textbf{Profit Rate ($R$)}} & \multicolumn{3}{c}{\textbf{TrueSkill ($S$)}} & \multicolumn{3}{c}{\textbf{Item Acquire Rate ($A$)}}\\
\cmidrule(lr){2-4} \cmidrule(lr){5-7} \cmidrule(lr){8-10}
\textbf{Master Agent's Setting} & \textbf{Master} & \textbf{Rival-1} & \textbf{Rival-2} & \textbf{Master} & \textbf{Rival-1} & \textbf{Rival-2} & \textbf{Master} & \textbf{Rival-1} & \textbf{Rival-2} \\
\midrule
Baseline w/o ToM & 23.45 & \textbf{24.08}  &  22.27  & \textbf{25.67}  & 25.00 &  24.11 &        0.77  & 0.80 &  \textbf{0.80} \\
ToM w/ True Persona & \textbf{26.18}   & 24.76 &  24.89  &   \textbf{25.00}   & 25.00 &    25.00 & 0.87 & 0.77 & \textbf{0.90}\\
ToM w/ True Persona + ExpAct &  \textcolor{red}{\textbf{35.81}}  & 21.32  & 22.03  &  \textcolor{red}{\textbf{30.34}} &  22.77 & 21.88 &    \textbf{0.87} & 0.73 & 0.77    \\
ToM w/ Inferred Pers. + ExpAct & \textbf{32.54}  &  21.14  & 22.08  & \textbf{28.56}  & 22.77    &  23.66 &   \textbf{1.00} & 0.80 & 0.83  \\
ToM w/ Inferred Pers. + RLAct & \textbf{30.01} & 21.55 & 23.61 & \textbf{27.23} & 23.22 & 24.55 & \textbf{0.93} & 0.87 & 0.90 \\
Second-Order ToM &  22.68 &  24.42 & \textbf{25.01} &   24.55   &  25.00   &   \textbf{25.45} & \textbf{0.93} & 0.90 & 0.87\\
\bottomrule
\end{tabular}
\end{small}
\vspace{-0.05in}
\caption{Equipping the Master with first-order ToM, allowing it to perceive competitors' true personas, increases its Profit Rate. Applying strategy further leads to a substantial improvement in both Profit Rate and TrueSkill score.}
\label{tab:experiment_strategy}
\vspace{-0.15in}
\end{table*}

Figure~\ref{fig:profile_persona} (left) shows the KL divergence scores, illustrating the Master Agent's ability to infer competitors' personas. As the number of matched items increases, the Master Agent becomes more accurate in profiling competitors, suggesting that effective inference depends on persona-driven behaviors. When no matched items are present, the lack of clear behavioral patterns leads to less accurate profiling.
Profiling a mix of two personas results in lower KL divergence than profiling a single persona. This is expected, as a balanced mix creates a more neutral agent—one that is less distinct and more evenly interested in different houses. As a result, its behavior is more predictable. In contrast, a single dominant persona leads to stronger preferences, making profiling more challenging. Figure~\ref{fig:profile_persona} (right) shows how the Master's profiling of a Rival Agent evolves after each bidding round.

\subsection{Profiling Capacity}
\label{sec:capacity}

The Master Agent's profiling ability diminishes as more bidders join. Increased interactions generate extensive bidding logs, making it harder for the Master to accurately infer competitor personas. 
In this experiment, we vary the number of competitors in an auction from \{2, 3, 4, 5, 6\}, with one Master Agent. Figure~\ref{fig:profile_persona} (middle) shows the Master's profiling results. We observe that the Master's profiling remains effective when there are fewer than five bidders. As more agents join, the Master agent struggles to infer personas accurately, as reflected by increasing KL divergence scores.

\subsection{Theory of Mind (ToM) Strategy}
\label{sec:4.5}

We examine whether the Master Agent can use ToM with persona knowledge from the profiling module (\S\ref{sec:persona}) to gain an advantage and whether the strategic module (\S\ref{sec:strategy}) enhances competitiveness. To ensure results are persona-independent, we rotate agent personas in a round-robin manner and report averaged results. Each agent has two items that align with their personas, making them the expected winners of those items while still aiming to maximize overall profit. 

Our ToM experiments consist of four settings: (a) \textbf{Baseline}: the Master operates without any ToM capabilities; (b) \textbf{ToM w/ True Persona}: the Master has the true personas of competing agents to assess whether persona knowledge enhances bidding performance, i.e., first-order ToM; (c) \textbf{ToM w/ Inferred Persona + ExpAct | RLAct}: instead of receiving true persona information, the Master infers competitors' personas using its profiling module and applies a strategic module; (d) \textbf{Second-Order ToM}: the Master is equipped with second-order ToM, allowing it to predict how rivals perceive its beliefs. All agents can infer other agents' personas and apply strategic modules. Details are provided in the Appendix~\ref{sec:2tom}.

Table~\ref{tab:experiment_strategy} presents the results of our ToM experiments. We observe that equipping the Master Agent with first-order ToM, allowing it to perceive competitors' true personas, increases its profit rate from 23.45\% (Baseline w/o ToM) to 26.18\%, while its TrueSkill remains at a similar level. When the agent relies solely on profiling without strategy, the profit increase is moderate. Applying strategy leads to a substantial improvement in both Profit Rate and TrueSkill score, boosting the profit rate from 26.18\% to 35.81\%. This emphasizes the need for expert strategy to help the Master Agent outperform competitors. When using inferred profiles instead of true personas, the Master experiences a modest performance gain, with its profit rate rising from 23.45\% to 32.54\%. This outcome aligns with expectations, suggesting that higher profitability requires more accurate profiling of competitors' preferences. Across all ToM settings, the Master demonstrates a higher item acquisition rate compared to the baseline, indicating the potential benefits of ToM in securing desired items.

\vspace{-0.05in}
\paragraph{ToM Strategy} Interestingly, using expert-crafted rules (ExpAct) as the strategy leads to a 32.54\% profit rate, outperforming the 30.01\% achieved by reinforcement learning (RLAct). RLAct most often chooses to bid on items that competitors are unlikely to target. This is likely because DPO training tends to guide the model toward benign strategies rather than aggressive or competitive ones. In other words, it avoids tactics that might hurt others to gain an edge. This indicates a potential limitation of reinforced fine-tuning in competitive settings.

\vspace{-0.05in}
\paragraph{Second-Order ToM} Master Agent with second-order ToM predicts how Rival Agents perceive its beliefs and persona. For example, if the Master Agent believes that Rival 2 has identified its home preference, it may strategically bid on other properties to mislead competitors. If rivals avoid bidding on its persona-aligned properties, the Master can secure them at lower prices.

Table~\ref{tab:experiment_strategy} shows that when all agents infer each other's beliefs, competition intensifies, making auctions more aggressive. While the Master Agent has a higher-order ToM, this does not always lead to higher profits. This suggests that ToM (understanding others' intentions) alone is not enough; it must be paired with expert strategies (taking the right actions accordingly) to maximize benefits.

\section{Conclusion}

\textsc{Harbor} provides a controlled environment for studying how LLM agents balance personal objectives with competitive strategy. Through extensive experiments, we demonstrate that persona-driven bidding significantly influences agent behavior. We show that LLM agents can infer competitors' personas with reasonable accuracy, though profiling effectiveness declines as the number of rivals increases. Our strategic module enhances decision-making by leveraging profiling insights.

\section*{Limitations}

Our testbed, \textsc{Harbor}, is a flexible tool for studying how AI agents interact in competitive auctions. While it helps uncover key insights about decision-making, there are limitations. \textsc{Harbor} uses LLM-based agents, which possess strong reasoning skills, and their decision-making is influenced by their training data and inherent biases. These agents can mimic strategic behaviors effectively, yet they may lack the creative problem-solving abilities that humans show in real-world auctions. This means their strategies might differ from those used by people in actual bidding scenarios. Our current evaluation metrics establish a foundation for measuring agent performance. Future versions of \textsc{Harbor} may incorporate additional features, such as modeling long-term trust between agents, simulating deception in negotiations, and addressing ethical considerations in competitive AI behavior.
These improvements will contribute to a comprehensive understanding of multi-agent interactions in competitive environments.

\bibliography{custom}

\begin{thebibliography}{104}
\providecommand{\natexlab}[1]{#1}

\bibitem[{Agashe et~al.(2025)Agashe, Han, Gan, Yang, Li, and Wang}]{agashe2025agent}
Saaket Agashe, Jiuzhou Han, Shuyu Gan, Jiachen Yang, Ang Li, and Xin~Eric Wang. 2025.
\newblock \href {https://openreview.net/forum?id=lIVRgt4nLv} {Agent s: An open agentic framework that uses computers like a human}.
\newblock In \emph{The Thirteenth International Conference on Learning Representations}.

\bibitem[{Amirizaniani et~al.(2024)Amirizaniani, Martin, Sivachenko, Mashhadi, and Shah}]{amirizaniani2024llms}
Maryam Amirizaniani, Elias Martin, Maryna Sivachenko, Afra Mashhadi, and Chirag Shah. 2024.
\newblock Do llms exhibit human-like reasoning? evaluating theory of mind in llms for open-ended responses.
\newblock \emph{arXiv preprint arXiv:2406.05659}.

\bibitem[{An et~al.(2024)An, Acquaye, Wang, Li, and Rudinger}]{an2024largelanguagemodelsdiscriminate}
Haozhe An, Christabel Acquaye, Colin Wang, Zongxia Li, and Rachel Rudinger. 2024.
\newblock \href {https://arxiv.org/abs/2406.10486} {Do large language models discriminate in hiring decisions on the basis of race, ethnicity, and gender?}
\newblock \emph{Preprint}, arXiv:2406.10486.

\bibitem[{Anthropic(2025)}]{anthropic_mcp}
Anthropic. 2025.
\newblock \href {https://docs.anthropic.com/en/docs/agents-and-tools/mcp} {Model context protocol (mcp)}.
\newblock Accessed: 2025-05-17.

\bibitem[{Anthropic(2024)}]{anthropic2024claude}
AI~Anthropic. 2024.
\newblock Claude 3.5 sonnet model card addendum.
\newblock \emph{Claude-3.5 Model Card}, 3(6).

\bibitem[{Asgari et~al.(2024)Asgari, Khani, and Khasahmadi}]{asgari2024mmlupro}
Saeid Asgari, Aliasghar Khani, and Amir~Hosein Khasahmadi. 2024.
\newblock \href {https://openreview.net/forum?id=3W2RQMdVXF} {{MMLU}-pro+: Evaluating higher-order reasoning and shortcut learning in {LLM}s}.
\newblock In \emph{Neurips Safe Generative AI Workshop 2024}.

\bibitem[{Bohnet et~al.(2024)Bohnet, Nova, Parisi, Swersky, Goshvadi, Dai, Schuurmans, Fiedel, and Sedghi}]{bohnet2024exploringbenchmarkingplanningcapabilities}
Bernd Bohnet, Azade Nova, Aaron~T Parisi, Kevin Swersky, Katayoon Goshvadi, Hanjun Dai, Dale Schuurmans, Noah Fiedel, and Hanie Sedghi. 2024.
\newblock \href {https://arxiv.org/abs/2406.13094} {Exploring and benchmarking the planning capabilities of large language models}.
\newblock \emph{Preprint}, arXiv:2406.13094.

\bibitem[{Chan et~al.(2024)Chan, Jiayang, Yim, Deng, Fan, Li, Liu, Zhang, Wang, and Song}]{chan-etal-2024-negotiationtom}
Chunkit Chan, Cheng Jiayang, Yauwai Yim, Zheye Deng, Wei Fan, Haoran Li, Xin Liu, Hongming Zhang, Weiqi Wang, and Yangqiu Song. 2024.
\newblock \href {https://doi.org/10.18653/v1/2024.findings-emnlp.244} {{N}egotiation{T}o{M}: A benchmark for stress-testing machine theory of mind on negotiation surrounding}.
\newblock In \emph{Findings of the Association for Computational Linguistics: EMNLP 2024}, pages 4211--4241, Miami, Florida, USA. Association for Computational Linguistics.

\bibitem[{Chen et~al.(2024{\natexlab{a}})Chen, Wang, Xu, Yuan, Zhang, Shi, Xie, Li, Yang, Zhu, Chen, Li, Chen, Hu, Wu, Ren, Fu, and Xiao}]{chen2024from}
Jiangjie Chen, Xintao Wang, Rui Xu, Siyu Yuan, Yikai Zhang, Wei Shi, Jian Xie, Shuang Li, Ruihan Yang, Tinghui Zhu, Aili Chen, Nianqi Li, Lida Chen, Caiyu Hu, Siye Wu, Scott Ren, Ziquan Fu, and Yanghua Xiao. 2024{\natexlab{a}}.
\newblock \href {https://openreview.net/forum?id=xrO70E8UIZ} {From persona to personalization: A survey on role-playing language agents}.
\newblock \emph{Transactions on Machine Learning Research}.
\newblock Survey Certification.

\bibitem[{Chen et~al.(2024{\natexlab{b}})Chen, Yuan, Ye, Majumder, and Richardson}]{chen2023put}
Jiangjie Chen, Siyu Yuan, Rong Ye, Bodhisattwa~Prasad Majumder, and Kyle Richardson. 2024{\natexlab{b}}.
\newblock Put your money where your mouth is: Evaluating strategic planning and execution of llm agents in an auction arena.
\newblock In \emph{NeurIPS 2024 Workshop on Open-World Agents}.

\bibitem[{Chen et~al.(2024{\natexlab{c}})Chen, Hu, Liu, Huang, Tu, He, and Wen}]{chen2024llmarenaassessingcapabilitieslarge}
Junzhe Chen, Xuming Hu, Shuodi Liu, Shiyu Huang, Wei-Wei Tu, Zhaofeng He, and Lijie Wen. 2024{\natexlab{c}}.
\newblock \href {https://arxiv.org/abs/2402.16499} {Llmarena: Assessing capabilities of large language models in dynamic multi-agent environments}.
\newblock \emph{Preprint}, arXiv:2402.16499.

\bibitem[{Chen et~al.(2024{\natexlab{d}})Chen, Zuo, Li, Zhang, Mei, and Bendersky}]{chen2024unlockingwhybuyingintroducing}
Tao Chen, Siqi Zuo, Cheng Li, Mingyang Zhang, Qiaozhu Mei, and Michael Bendersky. 2024{\natexlab{d}}.
\newblock \href {https://arxiv.org/abs/2402.13417} {Unlocking the `why' of buying: Introducing a new dataset and benchmark for purchase reason and post-purchase experience}.
\newblock \emph{Preprint}, arXiv:2402.13417.

\bibitem[{Chuang et~al.(2024)Chuang, Nirunwiroj, Studdiford, Goyal, Frigo, Yang, Shah, Hu, and Rogers}]{chuang-etal-2024-beyond}
Yun-Shiuan Chuang, Krirk Nirunwiroj, Zach Studdiford, Agam Goyal, Vincent~V. Frigo, Sijia Yang, Dhavan~V. Shah, Junjie Hu, and Timothy~T. Rogers. 2024.
\newblock \href {https://doi.org/10.18653/v1/2024.findings-emnlp.819} {Beyond demographics: Aligning role-playing {LLM}-based agents using human belief networks}.
\newblock In \emph{Findings of the Association for Computational Linguistics: EMNLP 2024}, pages 14010--14026, Miami, Florida, USA. Association for Computational Linguistics.

\bibitem[{Cross et~al.(2024)Cross, Xiang, Bhatia, Yamins, and Haber}]{cross2024hypothetical}
Logan Cross, Violet Xiang, Agam Bhatia, Daniel~LK Yamins, and Nick Haber. 2024.
\newblock Hypothetical minds: Scaffolding theory of mind for multi-agent tasks with large language models.
\newblock In \emph{NeurIPS 2024 Workshop on Open-World Agents}.

\bibitem[{Davidson et~al.(2024)Davidson, Veselovsky, Kosinski, and West}]{davidson2024evaluating}
Tim~Ruben Davidson, Veniamin Veselovsky, Michal Kosinski, and Robert West. 2024.
\newblock \href {https://openreview.net/forum?id=3ZqKxMHcAg} {Evaluating language model agency through negotiations}.
\newblock In \emph{The Twelfth International Conference on Learning Representations}.

\bibitem[{De~Weerd et~al.(2017)De~Weerd, Verbrugge, and Verheij}]{de2017negotiating}
Harmen De~Weerd, Rineke Verbrugge, and Bart Verheij. 2017.
\newblock Negotiating with other minds: the role of recursive theory of mind in negotiation with incomplete information.
\newblock \emph{Autonomous Agents and Multi-Agent Systems}, 31:250--287.

\bibitem[{Debenedetti et~al.(2024)Debenedetti, Zhang, Balunovic, Beurer-Kellner, Fischer, and Tram{\`e}r}]{debenedetti2024agentdojo}
Edoardo Debenedetti, Jie Zhang, Mislav Balunovic, Luca Beurer-Kellner, Marc Fischer, and Florian Tram{\`e}r. 2024.
\newblock \href {https://openreview.net/forum?id=m1YYAQjO3w} {Agentdojo: A dynamic environment to evaluate prompt injection attacks and defenses for {LLM} agents}.
\newblock In \emph{The Thirty-eight Conference on Neural Information Processing Systems Datasets and Benchmarks Track}.

\bibitem[{DeepSeek-AI et~al.(2025)DeepSeek-AI, Guo, Yang, Zhang, Song, Zhang, Xu, Zhu, Ma, Wang, Bi, Zhang, Yu, Wu, Wu, Gou, and etc}]{deepseekai2025deepseekr1incentivizingreasoningcapability}
DeepSeek-AI, Daya Guo, Dejian Yang, Haowei Zhang, Junxiao Song, Ruoyu Zhang, Runxin Xu, Qihao Zhu, Shirong Ma, Peiyi Wang, Xiao Bi, Xiaokang Zhang, Xingkai Yu, Yu~Wu, Z.~F. Wu, Zhibin Gou, and etc. 2025.
\newblock \href {https://arxiv.org/abs/2501.12948} {Deepseek-r1: Incentivizing reasoning capability in llms via reinforcement learning}.
\newblock \emph{Preprint}, arXiv:2501.12948.

\bibitem[{Dong et~al.(2024)Dong, Hu, and Collier}]{dong-etal-2024-llm}
Yijiang~River Dong, Tiancheng Hu, and Nigel Collier. 2024.
\newblock \href {https://doi.org/10.18653/v1/2024.findings-emnlp.592} {Can {LLM} be a personalized judge?}
\newblock In \emph{Findings of the Association for Computational Linguistics: EMNLP 2024}, pages 10126--10141, Miami, Florida, USA. Association for Computational Linguistics.

\bibitem[{Dutt et~al.(2021)Dutt, Sinha, Joshi, Chakraborty, Riggs, Yan, Bao, and Rose}]{dutt-etal-2021-resper}
Ritam Dutt, Sayan Sinha, Rishabh Joshi, Surya~Shekhar Chakraborty, Meredith Riggs, Xinru Yan, Haogang Bao, and Carolyn Rose. 2021.
\newblock \href {https://doi.org/10.18653/v1/2021.eacl-main.7} {{R}es{P}er: Computationally modelling resisting strategies in persuasive conversations}.
\newblock In \emph{Proceedings of the 16th Conference of the European Chapter of the Association for Computational Linguistics: Main Volume}, pages 78--90, Online. Association for Computational Linguistics.

\bibitem[{Eigner and Händler(2024)}]{eigner2024determinantsllmassisteddecisionmaking}
Eva Eigner and Thorsten Händler. 2024.
\newblock \href {https://arxiv.org/abs/2402.17385} {Determinants of llm-assisted decision-making}.
\newblock \emph{Preprint}, arXiv:2402.17385.

\bibitem[{Fan et~al.(2024)Fan, Chen, Jin, and He}]{caoyun2024canllm}
Caoyun Fan, Jindou Chen, Yaohui Jin, and Hao He. 2024.
\newblock Can large language models serve as rational players in game theory? a systematic analysis.
\newblock In \emph{Proceedings of the AAAI Conference on Artificial Intelligence}, volume~38, pages 17960--17967.

\bibitem[{Frisch and Giulianelli(2024)}]{frisch2024llmagentsinteractionmeasuring}
Ivar Frisch and Mario Giulianelli. 2024.
\newblock \href {https://arxiv.org/abs/2402.02896} {Llm agents in interaction: Measuring personality consistency and linguistic alignment in interacting populations of large language models}.
\newblock \emph{Preprint}, arXiv:2402.02896.

\bibitem[{Geng and Chang(2025)}]{geng2025realmbenchrealworldplanningbenchmark}
Longling Geng and Edward~Y. Chang. 2025.
\newblock \href {https://arxiv.org/abs/2502.18836} {Realm-bench: A real-world planning benchmark for llms and multi-agent systems}.
\newblock \emph{Preprint}, arXiv:2502.18836.

\bibitem[{Gottweis and Natarajan(2025)}]{gottweis2025ai}
Juraj Gottweis and Vivek Natarajan. 2025.
\newblock \href {https://research.google/blog/accelerating-scientific-breakthroughs-with-an-ai-co-scientist/} {Accelerating scientific breakthroughs with an ai co-scientist}.
\newblock Google Research Blog.

\bibitem[{Gu et~al.(2024)Gu, Zhu, Guo, Zhang, Cai, Shen, Chen, Ye, Dai, Gao et~al.}]{gu2024agent}
Zhouhong Gu, Xiaoxuan Zhu, Haoran Guo, Lin Zhang, Yin Cai, Hao Shen, Jiangjie Chen, Zheyu Ye, Yifei Dai, Yan Gao, et~al. 2024.
\newblock Agent group chat: An interactive group chat simulacra for better eliciting collective emergent behavior.
\newblock \emph{arXiv preprint arXiv:2403.13433}.

\bibitem[{Guo et~al.(2024)Guo, Chen, Wang, Chang, Pei, Chawla, Wiest, and Zhang}]{guo2024largelanguagemodelbased}
Taicheng Guo, Xiuying Chen, Yaqi Wang, Ruidi Chang, Shichao Pei, Nitesh~V. Chawla, Olaf Wiest, and Xiangliang Zhang. 2024.
\newblock \href {https://arxiv.org/abs/2402.01680} {Large language model based multi-agents: A survey of progress and challenges}.
\newblock \emph{Preprint}, arXiv:2402.01680.

\bibitem[{Gupta et~al.(2024)Gupta, Shrivastava, Deshpande, Kalyan, Clark, Sabharwal, and Khot}]{gupta2024bias}
Shashank Gupta, Vaishnavi Shrivastava, Ameet Deshpande, Ashwin Kalyan, Peter Clark, Ashish Sabharwal, and Tushar Khot. 2024.
\newblock \href {https://openreview.net/forum?id=kGteeZ18Ir} {Bias runs deep: Implicit reasoning biases in persona-assigned {LLM}s}.
\newblock In \emph{The Twelfth International Conference on Learning Representations}.

\bibitem[{He et~al.(2018)He, Chen, Balakrishnan, and Liang}]{he-etal-2018-decoupling}
He~He, Derek Chen, Anusha Balakrishnan, and Percy Liang. 2018.
\newblock \href {https://doi.org/10.18653/v1/D18-1256} {Decoupling strategy and generation in negotiation dialogues}.
\newblock In \emph{Proceedings of the 2018 Conference on Empirical Methods in Natural Language Processing}, pages 2333--2343, Brussels, Belgium. Association for Computational Linguistics.

\bibitem[{Herbrich et~al.(2007)Herbrich, Minka, and Graepel}]{herbrich2007trueskill}
Ralf Herbrich, Tom Minka, and Thore Graepel. 2007.
\newblock \href {https://www.microsoft.com/en-us/research/publication/trueskilltm-a-bayesian-skill-rating-system/} {Trueskill(tm): A bayesian skill rating system}.
\newblock In \emph{Advances in Neural Information Processing Systems 20}, pages 569--576. MIT Press.

\bibitem[{Herr et~al.(2024)Herr, Acero, Raileanu, P{\'e}rez-Ortiz, and Li}]{herr2024large}
Nathan Herr, Fernando Acero, Roberta Raileanu, Mar{\'\i}a P{\'e}rez-Ortiz, and Zhibin Li. 2024.
\newblock Are large language models strategic decision makers? a study of performance and bias in two-player non-zero-sum games.
\newblock \emph{arXiv preprint arXiv:2407.04467}.

\bibitem[{Hu and Collier(2024{\natexlab{a}})}]{hu-collier-2024-quantifying}
Tiancheng Hu and Nigel Collier. 2024{\natexlab{a}}.
\newblock \href {https://doi.org/10.18653/v1/2024.acl-long.554} {Quantifying the persona effect in {LLM} simulations}.
\newblock In \emph{Proceedings of the 62nd Annual Meeting of the Association for Computational Linguistics (Volume 1: Long Papers)}, pages 10289--10307, Bangkok, Thailand. Association for Computational Linguistics.

\bibitem[{Hu and Collier(2024{\natexlab{b}})}]{hu2024quantifying}
Tiancheng Hu and Nigel Collier. 2024{\natexlab{b}}.
\newblock Quantifying the persona effect in llm simulations.
\newblock \emph{arXiv preprint arXiv:2402.10811}.

\bibitem[{Hua et~al.(2024)Hua, Liu, Li, Amayuelas, Chen, Jiang, Jin, Fan, Sun, Wang, Wang, and Zhang}]{hua2024gametheoreticllmagentworkflow}
Wenyue Hua, Ollie Liu, Lingyao Li, Alfonso Amayuelas, Julie Chen, Lucas Jiang, Mingyu Jin, Lizhou Fan, Fei Sun, William Wang, Xintong Wang, and Yongfeng Zhang. 2024.
\newblock \href {https://arxiv.org/abs/2411.05990} {Game-theoretic llm: Agent workflow for negotiation games}.
\newblock \emph{Preprint}, arXiv:2411.05990.

\bibitem[{Huang et~al.(2024)Huang, Li, Lam, Liang, Wang, Yuan, Jiao, Wang, Tu, and Lyu}]{huang2024far}
Jen-tse Huang, Eric~John Li, Man~Ho Lam, Tian Liang, Wenxuan Wang, Youliang Yuan, Wenxiang Jiao, Xing Wang, Zhaopeng Tu, and Michael~R Lyu. 2024.
\newblock How far are we on the decision-making of llms? evaluating llms' gaming ability in multi-agent environments.
\newblock \emph{arXiv preprint arXiv:2403.11807}.

\bibitem[{Hurst et~al.(2024)Hurst, Lerer, Goucher, Perelman, Ramesh, Clark, Ostrow, Welihinda, Hayes, Radford et~al.}]{hurst2024gpt}
Aaron Hurst, Adam Lerer, Adam~P Goucher, Adam Perelman, Aditya Ramesh, Aidan Clark, AJ~Ostrow, Akila Welihinda, Alan Hayes, Alec Radford, et~al. 2024.
\newblock Gpt-4o system card.
\newblock \emph{arXiv preprint arXiv:2410.21276}.

\bibitem[{Kim et~al.(2024)Kim, Yang, and Jung}]{kim2024persona}
Junseok Kim, Nakyeong Yang, and Kyomin Jung. 2024.
\newblock Persona is a double-edged sword: Mitigating the negative impact of role-playing prompts in zero-shot reasoning tasks.
\newblock \emph{arXiv preprint arXiv:2408.08631}.

\bibitem[{Kullback and Leibler(1951)}]{kullback1951information}
Solomon Kullback and Richard~A Leibler. 1951.
\newblock On information and sufficiency.
\newblock \emph{The annals of mathematical statistics}, 22(1):79--86.

\bibitem[{Lee et~al.(2024)Lee, Lee, and Cho}]{lee2024language}
Bruce~W Lee, Yeongheon Lee, and Hyunsoo Cho. 2024.
\newblock Language models show stable value orientations across diverse role-plays.
\newblock \emph{arXiv preprint arXiv:2408.09049}.

\bibitem[{Leslie et~al.(2004)Leslie, Friedman, and German}]{leslie2004core}
Alan~M Leslie, Ori Friedman, and Tim~P German. 2004.
\newblock Core mechanisms in ‘theory of mind’.
\newblock \emph{Trends in cognitive sciences}, 8(12):528--533.

\bibitem[{Li et~al.(2025{\natexlab{a}})Li, Xie, Li, Tsung, Ding, and Li}]{li2025agentoriented}
Ao~Li, Yuexiang Xie, Songze Li, Fugee Tsung, Bolin Ding, and Yaliang Li. 2025{\natexlab{a}}.
\newblock \href {https://openreview.net/forum?id=EqcLAU6gyU} {Agent-oriented planning in multi-agent systems}.
\newblock In \emph{The Thirteenth International Conference on Learning Representations}.

\bibitem[{Li et~al.(2023{\natexlab{a}})Li, Hammoud, Itani, Khizbullin, and Ghanem}]{li2023camel}
Guohao Li, Hasan Hammoud, Hani Itani, Dmitrii Khizbullin, and Bernard Ghanem. 2023{\natexlab{a}}.
\newblock Camel: Communicative agents for" mind" exploration of large language model society.
\newblock \emph{Advances in Neural Information Processing Systems}, 36:51991--52008.

\bibitem[{Li et~al.(2025{\natexlab{b}})Li, Yang, Zhang, Deng, Wang, and Chua}]{li2025helloagainllmpoweredpersonalized}
Hao Li, Chenghao Yang, An~Zhang, Yang Deng, Xiang Wang, and Tat-Seng Chua. 2025{\natexlab{b}}.
\newblock \href {https://arxiv.org/abs/2406.05925} {Hello again! llm-powered personalized agent for long-term dialogue}.
\newblock \emph{Preprint}, arXiv:2406.05925.

\bibitem[{Li et~al.(2025{\natexlab{c}})Li, Yang, Zhang, Deng, Wang, and Chua}]{li-etal-2025-hello}
Hao Li, Chenghao Yang, An~Zhang, Yang Deng, Xiang Wang, and Tat-Seng Chua. 2025{\natexlab{c}}.
\newblock \href {https://aclanthology.org/2025.naacl-long.272/} {Hello again! {LLM}-powered personalized agent for long-term dialogue}.
\newblock In \emph{Proceedings of the 2025 Conference of the Nations of the Americas Chapter of the Association for Computational Linguistics: Human Language Technologies (Volume 1: Long Papers)}, pages 5259--5276, Albuquerque, New Mexico. Association for Computational Linguistics.

\bibitem[{Li et~al.(2024{\natexlab{a}})Li, Chen, Zhang, and Liu}]{li2024laspsurveyingstateoftheartlarge}
Haoming Li, Zhaoliang Chen, Jonathan Zhang, and Fei Liu. 2024{\natexlab{a}}.
\newblock \href {https://arxiv.org/abs/2409.01806} {Lasp: Surveying the state-of-the-art in large language model-assisted ai planning}.
\newblock \emph{Preprint}, arXiv:2409.01806.

\bibitem[{Li et~al.(2023{\natexlab{b}})Li, Chong, Stepputtis, Campbell, Hughes, Lewis, and Sycara}]{Li_2023}
Huao Li, Yu~Chong, Simon Stepputtis, Joseph Campbell, Dana Hughes, Charles Lewis, and Katia Sycara. 2023{\natexlab{b}}.
\newblock \href {https://doi.org/10.18653/v1/2023.emnlp-main.13} {Theory of mind for multi-agent collaboration via large language models}.
\newblock In \emph{Proceedings of the 2023 Conference on Empirical Methods in Natural Language Processing}. Association for Computational Linguistics.

\bibitem[{Li et~al.(2025{\natexlab{d}})Li, Zhao, Wang, Wang, Zhou, Srivastava, Gokmen, Lee, Li, Zhang, Liu, Liang, Fei-Fei, Mao, and Wu}]{li2025embodiedagentinterfacebenchmarking}
Manling Li, Shiyu Zhao, Qineng Wang, Kangrui Wang, Yu~Zhou, Sanjana Srivastava, Cem Gokmen, Tony Lee, Li~Erran Li, Ruohan Zhang, Weiyu Liu, Percy Liang, Li~Fei-Fei, Jiayuan Mao, and Jiajun Wu. 2025{\natexlab{d}}.
\newblock \href {https://arxiv.org/abs/2410.07166} {Embodied agent interface: Benchmarking llms for embodied decision making}.
\newblock \emph{Preprint}, arXiv:2410.07166.

\bibitem[{Li et~al.(2024{\natexlab{b}})Li, Zang, Ma, Guo, Zheng, Liu, Niu, Wang, Yang, Liu, Zhong, Zhou, Huang, and Zhang}]{li2024autokagglemultiagentframeworkautonomous}
Ziming Li, Qianbo Zang, David Ma, Jiawei Guo, Tuney Zheng, Minghao Liu, Xinyao Niu, Yue Wang, Jian Yang, Jiaheng Liu, Wanjun Zhong, Wangchunshu Zhou, Wenhao Huang, and Ge~Zhang. 2024{\natexlab{b}}.
\newblock \href {https://arxiv.org/abs/2410.20424} {Autokaggle: A multi-agent framework for autonomous data science competitions}.
\newblock \emph{Preprint}, arXiv:2410.20424.

\bibitem[{Liao et~al.(2025)Liao, Wen, Wang, and Zhang}]{liao2025marftmultiagentreinforcementfinetuning}
Junwei Liao, Muning Wen, Jun Wang, and Weinan Zhang. 2025.
\newblock \href {https://arxiv.org/abs/2504.16129} {Marft: Multi-agent reinforcement fine-tuning}.
\newblock \emph{Preprint}, arXiv:2504.16129.

\bibitem[{Light et~al.(2023)Light, Cai, Shen, and Hu}]{light2023avalonbench}
Jonathan Light, Min Cai, Sheng Shen, and Ziniu Hu. 2023.
\newblock Avalonbench: Evaluating llms playing the game of avalon.
\newblock In \emph{NeurIPS 2023 Foundation Models for Decision Making Workshop}.

\bibitem[{Lin et~al.(2024)Lin, Tomlin, Andreas, and Eisner}]{lin-etal-2024-decision}
Jessy Lin, Nicholas Tomlin, Jacob Andreas, and Jason Eisner. 2024.
\newblock \href {https://doi.org/10.1162/tacl_a_00679} {Decision-oriented dialogue for human-{AI} collaboration}.
\newblock \emph{Transactions of the Association for Computational Linguistics}, 12:892--911.

\bibitem[{Liu et~al.(2024{\natexlab{a}})Liu, Fu, Yogatama, and Neiswanger}]{liu2024dellma}
Ollie Liu, Deqing Fu, Dani Yogatama, and Willie Neiswanger. 2024{\natexlab{a}}.
\newblock \href {https://arxiv.org/abs/2402.02392} {{D}e{L}{L}{M}a: {D}ecision {M}aking {U}nder {U}ncertainty with {L}arge {L}anguage {M}odels}.
\newblock \emph{Preprint}, arXiv:2402.02392.

\bibitem[{Liu et~al.(2024{\natexlab{b}})Liu, Geng, Peterson, Sucholutsky, and Griffiths}]{liu2024large}
Ryan Liu, Jiayi Geng, Joshua~C Peterson, Ilia Sucholutsky, and Thomas~L Griffiths. 2024{\natexlab{b}}.
\newblock Large language models assume people are more rational than we really are.
\newblock \emph{arXiv preprint arXiv:2406.17055}.

\bibitem[{Liu et~al.(2024{\natexlab{c}})Liu, Yu, Zhang, Xu, Lei, Lai, Gu, Ding, Men, Yang, Zhang, Deng, Zeng, Du, Zhang, Shen, Zhang, Su, Sun, Huang, Dong, and Tang}]{liu2024agentbench}
Xiao Liu, Hao Yu, Hanchen Zhang, Yifan Xu, Xuanyu Lei, Hanyu Lai, Yu~Gu, Hangliang Ding, Kaiwen Men, Kejuan Yang, Shudan Zhang, Xiang Deng, Aohan Zeng, Zhengxiao Du, Chenhui Zhang, Sheng Shen, Tianjun Zhang, Yu~Su, Huan Sun, Minlie Huang, Yuxiao Dong, and Jie Tang. 2024{\natexlab{c}}.
\newblock \href {https://openreview.net/forum?id=zAdUB0aCTQ} {Agentbench: Evaluating {LLM}s as agents}.
\newblock In \emph{The Twelfth International Conference on Learning Representations}.

\bibitem[{Lu et~al.(2024)Lu, Hu, Foroosh, Jin, and Liu}]{lu2024struxllmdecisionmakingstructured}
Yiming Lu, Yebowen Hu, Hassan Foroosh, Wei Jin, and Fei Liu. 2024.
\newblock \href {https://arxiv.org/abs/2410.12583} {Strux: An llm for decision-making with structured explanations}.
\newblock \emph{Preprint}, arXiv:2410.12583.

\bibitem[{Meulemans et~al.(2025)Meulemans, Kobayashi, von Oswald, Scherrer, Elmoznino, Richards, Lajoie, y~Arcas, and Sacramento}]{meulemans2025multiagent}
Alexander Meulemans, Seijin Kobayashi, Johannes von Oswald, Nino Scherrer, Eric Elmoznino, Blake~Aaron Richards, Guillaume Lajoie, Blaise~Aguera y~Arcas, and Joao Sacramento. 2025.
\newblock \href {https://openreview.net/forum?id=GkWA6NjePN} {Multi-agent cooperation through learning-aware policy gradients}.
\newblock In \emph{The Thirteenth International Conference on Learning Representations}.

\bibitem[{{Microsoft}(2025)}]{microsoft_autogen}
{Microsoft}. 2025.
\newblock {AutoGen: A Open-Source Framework for Agentic AI}.
\newblock \url{https://github.com/microsoft/autogen}.
\newblock Accessed: 2025-05-17.

\bibitem[{{OpenAI}(2025)}]{openai_agents_sdk}
{OpenAI}. 2025.
\newblock \href {https://openai.github.io/openai-agents-python/} {Openai agents sdk documentation}.
\newblock Accessed: 2025-05-17.

\bibitem[{Park et~al.(2023)Park, O'Brien, Cai, Morris, Liang, and Bernstein}]{park2023generative}
Joon~Sung Park, Joseph O'Brien, Carrie~Jun Cai, Meredith~Ringel Morris, Percy Liang, and Michael~S Bernstein. 2023.
\newblock Generative agents: Interactive simulacra of human behavior.
\newblock In \emph{Proceedings of the 36th annual acm symposium on user interface software and technology}, pages 1--22.

\bibitem[{Qin et~al.(2023)Qin, Liang, Ye, Zhu, Yan, Lu, Lin, Cong, Tang, Qian, Zhao, Hong, Tian, Xie, Zhou, Gerstein, Li, Liu, and Sun}]{qin2023toolllmfacilitatinglargelanguage}
Yujia Qin, Shihao Liang, Yining Ye, Kunlun Zhu, Lan Yan, Yaxi Lu, Yankai Lin, Xin Cong, Xiangru Tang, Bill Qian, Sihan Zhao, Lauren Hong, Runchu Tian, Ruobing Xie, Jie Zhou, Mark Gerstein, Dahai Li, Zhiyuan Liu, and Maosong Sun. 2023.
\newblock \href {https://arxiv.org/abs/2307.16789} {Toolllm: Facilitating large language models to master 16000+ real-world apis}.
\newblock \emph{Preprint}, arXiv:2307.16789.

\bibitem[{Rafailov et~al.(2024)Rafailov, Sharma, Mitchell, Ermon, Manning, and Finn}]{rafailov2024directpreferenceoptimizationlanguage}
Rafael Rafailov, Archit Sharma, Eric Mitchell, Stefano Ermon, Christopher~D. Manning, and Chelsea Finn. 2024.
\newblock \href {https://arxiv.org/abs/2305.18290} {Direct preference optimization: Your language model is secretly a reward model}.
\newblock \emph{Preprint}, arXiv:2305.18290.

\bibitem[{Saha et~al.(2023)Saha, Hase, and Bansal}]{saha2023can}
Swarnadeep Saha, Peter Hase, and Mohit Bansal. 2023.
\newblock \href {https://openreview.net/forum?id=IacxcFpvWQ} {Can language models teach? teacher explanations improve student performance via personalization}.
\newblock In \emph{Thirty-seventh Conference on Neural Information Processing Systems}.

\bibitem[{Samuel et~al.(2024)Samuel, Zou, Zhou, Chaudhari, Kalyan, Rajpurohit, Deshpande, Narasimhan, and Murahari}]{samuel2024personagymevaluatingpersonaagents}
Vinay Samuel, Henry~Peng Zou, Yue Zhou, Shreyas Chaudhari, Ashwin Kalyan, Tanmay Rajpurohit, Ameet Deshpande, Karthik Narasimhan, and Vishvak Murahari. 2024.
\newblock \href {https://arxiv.org/abs/2407.18416} {Personagym: Evaluating persona agents and llms}.
\newblock \emph{Preprint}, arXiv:2407.18416.

\bibitem[{Sap et~al.(2022)Sap, Le~Bras, Fried, and Choi}]{sap-etal-2022-neural}
Maarten Sap, Ronan Le~Bras, Daniel Fried, and Yejin Choi. 2022.
\newblock \href {https://doi.org/10.18653/v1/2022.emnlp-main.248} {Neural theory-of-mind? on the limits of social intelligence in large {LM}s}.
\newblock In \emph{Proceedings of the 2022 Conference on Empirical Methods in Natural Language Processing}, pages 3762--3780, Abu Dhabi, United Arab Emirates. Association for Computational Linguistics.

\bibitem[{Shinn et~al.(2024)Shinn, Cassano, Gopinath, Narasimhan, and Yao}]{shinn2024reflexion}
Noah Shinn, Federico Cassano, Ashwin Gopinath, Karthik Narasimhan, and Shunyu Yao. 2024.
\newblock Reflexion: Language agents with verbal reinforcement learning.
\newblock \emph{Advances in Neural Information Processing Systems}, 36.

\bibitem[{Skrynnik et~al.(2025)Skrynnik, Andreychuk, Borzilov, Chernyavskiy, Yakovlev, and Panov}]{skrynnik2025pogema}
Alexey Skrynnik, Anton Andreychuk, Anatolii Borzilov, Alexander Chernyavskiy, Konstantin Yakovlev, and Aleksandr Panov. 2025.
\newblock \href {https://openreview.net/forum?id=6VgwE2tCRm} {{POGEMA}: A benchmark platform for cooperative multi-agent pathfinding}.
\newblock In \emph{The Thirteenth International Conference on Learning Representations}.

\bibitem[{Street(2024)}]{street2024llm}
Winnie Street. 2024.
\newblock Llm theory of mind and alignment: Opportunities and risks.
\newblock \emph{arXiv preprint arXiv:2405.08154}.

\bibitem[{Sun et~al.(2024)Sun, Zhan, and Such}]{sun2024building}
Guangzhi Sun, Xiao Zhan, and Jose Such. 2024.
\newblock Building better ai agents: A provocation on the utilisation of persona in llm-based conversational agents.
\newblock In \emph{Proceedings of the 6th ACM Conference on Conversational User Interfaces}, pages 1--6.

\bibitem[{Surapaneni et~al.(2025)Surapaneni, Jha, Vakoc, and Segal}]{google_a2a_2025}
Rao Surapaneni, Miku Jha, Michael Vakoc, and Todd Segal. 2025.
\newblock \href {https://developers.googleblog.com/en/a2a-a-new-era-of-agent-interoperability/} {Announcing the agent2agent protocol (a2a)}.
\newblock Accessed: 2025-05-17.

\bibitem[{Tan and Lee(2025)}]{tan-lee-2025-unmasking}
Bryan Chen~Zhengyu Tan and Roy Ka-Wei Lee. 2025.
\newblock \href {https://aclanthology.org/2025.naacl-long.50/} {Unmasking implicit bias: Evaluating persona-prompted {LLM} responses in power-disparate social scenarios}.
\newblock In \emph{Proceedings of the 2025 Conference of the Nations of the Americas Chapter of the Association for Computational Linguistics: Human Language Technologies (Volume 1: Long Papers)}, pages 1075--1108, Albuquerque, New Mexico. Association for Computational Linguistics.

\bibitem[{Tran et~al.(2025)Tran, Dao, Nguyen, Pham, O'Sullivan, and Nguyen}]{tran2025multiagentcollaborationmechanismssurvey}
Khanh-Tung Tran, Dung Dao, Minh-Duong Nguyen, Quoc-Viet Pham, Barry O'Sullivan, and Hoang~D. Nguyen. 2025.
\newblock \href {https://arxiv.org/abs/2501.06322} {Multi-agent collaboration mechanisms: A survey of llms}.
\newblock \emph{Preprint}, arXiv:2501.06322.

\bibitem[{Tseng et~al.(2024)Tseng, Huang, Hsiao, Chen, Huang, Meng, and Chen}]{tseng-etal-2024-two}
Yu-Min Tseng, Yu-Chao Huang, Teng-Yun Hsiao, Wei-Lin Chen, Chao-Wei Huang, Yu~Meng, and Yun-Nung Chen. 2024.
\newblock \href {https://doi.org/10.18653/v1/2024.findings-emnlp.969} {Two tales of persona in {LLM}s: A survey of role-playing and personalization}.
\newblock In \emph{Findings of the Association for Computational Linguistics: EMNLP 2024}, pages 16612--16631, Miami, Florida, USA. Association for Computational Linguistics.

\bibitem[{Valmeekam et~al.(2023)Valmeekam, Marquez, Sreedharan, and Kambhampati}]{valmeekam2023planningabilitieslargelanguage}
Karthik Valmeekam, Matthew Marquez, Sarath Sreedharan, and Subbarao Kambhampati. 2023.
\newblock \href {https://arxiv.org/abs/2305.15771} {On the planning abilities of large language models : A critical investigation}.
\newblock \emph{Preprint}, arXiv:2305.15771.

\bibitem[{Valmeekam et~al.(2024)Valmeekam, Stechly, Gundawar, and Kambhampati}]{valmeekam2024planningstrawberryfieldsevaluating}
Karthik Valmeekam, Kaya Stechly, Atharva Gundawar, and Subbarao Kambhampati. 2024.
\newblock \href {https://arxiv.org/abs/2410.02162} {Planning in strawberry fields: Evaluating and improving the planning and scheduling capabilities of lrm o1}.
\newblock \emph{Preprint}, arXiv:2410.02162.

\bibitem[{van Duijn et~al.(2023)van Duijn, van Dijk, Kouwenhoven, de~Valk, Spruit, and van~der Putten}]{van-duijn-etal-2023-theory}
Max van Duijn, Bram van Dijk, Tom Kouwenhoven, Werner de~Valk, Marco Spruit, and Peter van~der Putten. 2023.
\newblock \href {https://doi.org/10.18653/v1/2023.conll-1.25} {Theory of mind in large language models: Examining performance of 11 state-of-the-art models vs. children aged 7-10 on advanced tests}.
\newblock In \emph{Proceedings of the 27th Conference on Computational Natural Language Learning (CoNLL)}, pages 389--402, Singapore. Association for Computational Linguistics.

\bibitem[{Wang et~al.(2024{\natexlab{a}})Wang, Yu, Zhang, Qi, Sap, Bisk, Neubig, and Zhu}]{wang-etal-2024-sotopia}
Ruiyi Wang, Haofei Yu, Wenxin Zhang, Zhengyang Qi, Maarten Sap, Yonatan Bisk, Graham Neubig, and Hao Zhu. 2024{\natexlab{a}}.
\newblock \href {https://doi.org/10.18653/v1/2024.acl-long.698} {{SOTOPIA}-{\ensuremath{\pi}}: Interactive learning of socially intelligent language agents}.
\newblock In \emph{Proceedings of the 62nd Annual Meeting of the Association for Computational Linguistics (Volume 1: Long Papers)}, pages 12912--12940, Bangkok, Thailand. Association for Computational Linguistics.

\bibitem[{Wang et~al.(2024{\natexlab{b}})Wang, Zhang, Feng, Wang, and Tang}]{wang2024battleagentbenchbenchmarkevaluatingcooperation}
Wei Wang, Dan Zhang, Tao Feng, Boyan Wang, and Jie Tang. 2024{\natexlab{b}}.
\newblock \href {https://arxiv.org/abs/2408.15971} {Battleagentbench: A benchmark for evaluating cooperation and competition capabilities of language models in multi-agent systems}.
\newblock \emph{Preprint}, arXiv:2408.15971.

\bibitem[{Wang et~al.(2024{\natexlab{c}})Wang, Chen, Yuan, Zhang, Li, Peng, and Ji}]{wang2024executable}
Xingyao Wang, Yangyi Chen, Lifan Yuan, Yizhe Zhang, Yunzhu Li, Hao Peng, and Heng Ji. 2024{\natexlab{c}}.
\newblock \href {https://openreview.net/forum?id=8oJyuXfrPv} {Executable code actions elicit better {LLM} agents}.
\newblock In \emph{ICLR 2024 Workshop on Large Language Model (LLM) Agents}.

\bibitem[{Wang et~al.(2024{\natexlab{d}})Wang, Mao, Wu, Ge, Wei, and Ji}]{wang2024unleashingemergentcognitivesynergy}
Zhenhailong Wang, Shaoguang Mao, Wenshan Wu, Tao Ge, Furu Wei, and Heng Ji. 2024{\natexlab{d}}.
\newblock \href {https://arxiv.org/abs/2307.05300} {Unleashing the emergent cognitive synergy in large language models: A task-solving agent through multi-persona self-collaboration}.
\newblock \emph{Preprint}, arXiv:2307.05300.

\bibitem[{Wei et~al.(2022)Wei, Wang, Schuurmans, Bosma, brian ichter, Xia, Chi, Le, and Zhou}]{wei2022chain}
Jason Wei, Xuezhi Wang, Dale Schuurmans, Maarten Bosma, brian ichter, Fei Xia, Ed~H. Chi, Quoc~V Le, and Denny Zhou. 2022.
\newblock \href {https://openreview.net/forum?id=_VjQlMeSB_J} {Chain of thought prompting elicits reasoning in large language models}.
\newblock In \emph{Advances in Neural Information Processing Systems}.

\bibitem[{Weissburg et~al.(2025)Weissburg, Anand, Levy, and Jeong}]{weissburg-etal-2025-llms}
Iain Weissburg, Sathvika Anand, Sharon Levy, and Haewon Jeong. 2025.
\newblock \href {https://aclanthology.org/2025.findings-naacl.314/} {{LLM}s are biased teachers: Evaluating {LLM} bias in personalized education}.
\newblock In \emph{Findings of the Association for Computational Linguistics: NAACL 2025}, pages 5650--5698, Albuquerque, New Mexico. Association for Computational Linguistics.

\bibitem[{Wu et~al.(2024{\natexlab{a}})Wu, Shi, Sun, and Liu}]{wu-etal-2024-deciphering}
Dekun Wu, Haochen Shi, Zhiyuan Sun, and Bang Liu. 2024{\natexlab{a}}.
\newblock \href {https://doi.org/10.18653/v1/2024.findings-acl.490} {Deciphering digital detectives: Understanding {LLM} behaviors and capabilities in multi-agent mystery games}.
\newblock In \emph{Findings of the Association for Computational Linguistics: ACL 2024}, pages 8225--8291, Bangkok, Thailand. Association for Computational Linguistics.

\bibitem[{Wu et~al.(2024{\natexlab{b}})Wu, Bansal, Zhang, Wu, Li, Zhu, Jiang, Zhang, Zhang, Liu, Awadallah, White, Burger, and Wang}]{wu2024autogen}
Qingyun Wu, Gagan Bansal, Jieyu Zhang, Yiran Wu, Beibin Li, Erkang Zhu, Li~Jiang, Xiaoyun Zhang, Shaokun Zhang, Jiale Liu, Ahmed~Hassan Awadallah, Ryen~W White, Doug Burger, and Chi Wang. 2024{\natexlab{b}}.
\newblock \href {https://openreview.net/forum?id=tEAF9LBdgu} {Autogen: Enabling next-gen {LLM} applications via multi-agent conversation}.

\bibitem[{Wu et~al.(2024{\natexlab{c}})Wu, Peng, Zheng, Liu, Han, Kwon, Onizuka, Tang, and Xiao}]{wu-etal-2024-shall}
Zengqing Wu, Run Peng, Shuyuan Zheng, Qianying Liu, Xu~Han, Brian~I. Kwon, Makoto Onizuka, Shaojie Tang, and Chuan Xiao. 2024{\natexlab{c}}.
\newblock \href {https://doi.org/10.18653/v1/2024.findings-emnlp.297} {Shall we team up: Exploring spontaneous cooperation of competing {LLM} agents}.
\newblock In \emph{Findings of the Association for Computational Linguistics: EMNLP 2024}, pages 5163--5186, Miami, Florida, USA. Association for Computational Linguistics.

\bibitem[{Xie et~al.(2024{\natexlab{a}})Xie, Chen, Jia, Ye, Lai, Shu, Gu, Bibi, Hu, Jurgens et~al.}]{xie2024can}
Chengxing Xie, Canyu Chen, Feiran Jia, Ziyu Ye, Shiyang Lai, Kai Shu, Jindong Gu, Adel Bibi, Ziniu Hu, David Jurgens, et~al. 2024{\natexlab{a}}.
\newblock Can large language model agents simulate human trust behavior?
\newblock In \emph{The Thirty-eighth Annual Conference on Neural Information Processing Systems}.

\bibitem[{Xie et~al.(2024{\natexlab{b}})Xie, Zhang, Chen, Li, Zhao, Cao, Hua, Cheng, Shin, Lei, Liu, Xu, Zhou, Savarese, Xiong, Zhong, and Yu}]{xie2024osworldbenchmarkingmultimodalagents}
Tianbao Xie, Danyang Zhang, Jixuan Chen, Xiaochuan Li, Siheng Zhao, Ruisheng Cao, Toh~Jing Hua, Zhoujun Cheng, Dongchan Shin, Fangyu Lei, Yitao Liu, Yiheng Xu, Shuyan Zhou, Silvio Savarese, Caiming Xiong, Victor Zhong, and Tao Yu. 2024{\natexlab{b}}.
\newblock \href {https://arxiv.org/abs/2404.07972} {Osworld: Benchmarking multimodal agents for open-ended tasks in real computer environments}.
\newblock \emph{Preprint}, arXiv:2404.07972.

\bibitem[{Xu et~al.(2024{\natexlab{a}})Xu, Song, Li, Tang, Jain, Bao, Wang, Zhou, Guo, Cao, Yang, Lu, Martin, Su, Maben, Mehta, Chi, Jang, Xie, Zhou, and Neubig}]{xu2024theagentcompanybenchmarkingllmagents}
Frank~F. Xu, Yufan Song, Boxuan Li, Yuxuan Tang, Kritanjali Jain, Mengxue Bao, Zora~Z. Wang, Xuhui Zhou, Zhitong Guo, Murong Cao, Mingyang Yang, Hao~Yang Lu, Amaad Martin, Zhe Su, Leander Maben, Raj Mehta, Wayne Chi, Lawrence Jang, Yiqing Xie, Shuyan Zhou, and Graham Neubig. 2024{\natexlab{a}}.
\newblock \href {https://arxiv.org/abs/2412.14161} {Theagentcompany: Benchmarking llm agents on consequential real world tasks}.
\newblock \emph{Preprint}, arXiv:2412.14161.

\bibitem[{Xu et~al.(2024{\natexlab{b}})Xu, Wang, Chen, Yuan, Yuan, Liang, Chen, Dong, and Xiao}]{xu2024character}
Rui Xu, Xintao Wang, Jiangjie Chen, Siyu Yuan, Xinfeng Yuan, Jiaqing Liang, Zulong Chen, Xiaoqing Dong, and Yanghua Xiao. 2024{\natexlab{b}}.
\newblock Character is destiny: Can large language models simulate persona-driven decisions in role-playing?
\newblock \emph{arXiv preprint arXiv:2404.12138}.

\bibitem[{Yang et~al.(2024{\natexlab{a}})Yang, Korecki, Dailisan, Hausladen, and Helbing}]{yang2024llm}
Joshua~C Yang, Marcin Korecki, Damian Dailisan, Carina~I Hausladen, and Dirk Helbing. 2024{\natexlab{a}}.
\newblock Llm voting: Human choices and ai collective decision making.
\newblock \emph{arXiv preprint arXiv:2402.01766}.

\bibitem[{Yang et~al.(2024{\natexlab{b}})Yang, Li, Wen, Peng, Tang, and Liu}]{yang2024}
Kaiqi Yang, Hang Li, Hongzhi Wen, Tai-Quan Peng, Jiliang Tang, and Hui Liu. 2024{\natexlab{b}}.
\newblock \href {https://doi.org/10.18653/v1/2024.findings-emnlp.153} {Are large language models ({LLM}s) good social predictors?}
\newblock In \emph{Findings of the Association for Computational Linguistics: EMNLP 2024}, pages 2718--2730, Miami, Florida, USA. Association for Computational Linguistics.

\bibitem[{Yang et~al.(2024{\natexlab{c}})Yang, Chen, Zhang, Yuan, Chen, Richardson, Xiao, and Yang}]{yang2024selfgoal}
Ruihan Yang, Jiangjie Chen, Yikai Zhang, Siyu Yuan, Aili Chen, Kyle Richardson, Yanghua Xiao, and Deqing Yang. 2024{\natexlab{c}}.
\newblock Selfgoal: Your language agents already know how to achieve high-level goals.
\newblock \emph{arXiv preprint arXiv:2406.04784}.

\bibitem[{Yang et~al.(2021)Yang, Chen, and Narasimhan}]{yang-etal-2021-improving}
Runzhe Yang, Jingxiao Chen, and Karthik Narasimhan. 2021.
\newblock \href {https://doi.org/10.18653/v1/2021.acl-long.56} {Improving dialog systems for negotiation with personality modeling}.
\newblock In \emph{Proceedings of the 59th Annual Meeting of the Association for Computational Linguistics and the 11th International Joint Conference on Natural Language Processing (Volume 1: Long Papers)}, pages 681--693, Online. Association for Computational Linguistics.

\bibitem[{Yao et~al.(2025)Yao, Cheng, Wei, and Xu}]{anonymous2025singleagent}
Fan Yao, Yuwei Cheng, Ermin Wei, and Haifeng Xu. 2025.
\newblock \href {https://openreview.net/forum?id=46xYl55hdc} {Single-agent poisoning attacks suffice to ruin multi-agent learning}.
\newblock In \emph{The Thirteenth International Conference on Learning Representations}.

\bibitem[{Yao et~al.(2023)Yao, Zhao, Yu, Du, Shafran, Narasimhan, and Cao}]{yao2023reactsynergizingreasoningacting}
Shunyu Yao, Jeffrey Zhao, Dian Yu, Nan Du, Izhak Shafran, Karthik Narasimhan, and Yuan Cao. 2023.
\newblock \href {https://arxiv.org/abs/2210.03629} {React: Synergizing reasoning and acting in language models}.
\newblock \emph{Preprint}, arXiv:2210.03629.

\bibitem[{Ye et~al.(2024)Ye, Wang, Huang, Chen, Zhang, Moniz, Gao, Geyer, Huang, Chen et~al.}]{ye2024justice}
Jiayi Ye, Yanbo Wang, Yue Huang, Dongping Chen, Qihui Zhang, Nuno Moniz, Tian Gao, Werner Geyer, Chao Huang, Pin-Yu Chen, et~al. 2024.
\newblock Justice or prejudice? quantifying biases in llm-as-a-judge.
\newblock \emph{arXiv preprint arXiv:2410.02736}.

\bibitem[{Yim et~al.(2024)Yim, Chan, Shi, Deng, Fan, Zheng, and Song}]{yim2024evaluating}
Yauwai Yim, Chunkit Chan, Tianyu Shi, Zheye Deng, Wei Fan, Tianshi Zheng, and Yangqiu Song. 2024.
\newblock Evaluating and enhancing llms agent based on theory of mind in guandan: A multi-player cooperative game under imperfect information.
\newblock \emph{arXiv preprint arXiv:2408.02559}.

\bibitem[{Zhang et~al.(2024{\natexlab{a}})Zhang, Xu, Zhang, Liu, Hooi, and Deng}]{zhang-etal-2024-exploring}
Jintian Zhang, Xin Xu, Ningyu Zhang, Ruibo Liu, Bryan Hooi, and Shumin Deng. 2024{\natexlab{a}}.
\newblock \href {https://doi.org/10.18653/v1/2024.acl-long.782} {Exploring collaboration mechanisms for {LLM} agents: A social psychology view}.
\newblock In \emph{Proceedings of the 62nd Annual Meeting of the Association for Computational Linguistics (Volume 1: Long Papers)}, pages 14544--14607, Bangkok, Thailand. Association for Computational Linguistics.

\bibitem[{Zhang et~al.(2024{\natexlab{b}})Zhang, Wang, Zhang, Chen, Gao, Wang, Zhang, Wang, and Wen}]{zhang2024mutual}
Shao Zhang, Xihuai Wang, Wenhao Zhang, Yongshan Chen, Landi Gao, Dakuo Wang, Weinan Zhang, Xinbing Wang, and Ying Wen. 2024{\natexlab{b}}.
\newblock Mutual theory of mind in human-ai collaboration: An empirical study with llm-driven ai agents in a real-time shared workspace task.
\newblock \emph{arXiv preprint arXiv:2409.08811}.

\bibitem[{Zhang et~al.(2024{\natexlab{c}})Zhang, Mao, Ge, Wang, de~Wynter, Xia, Wu, Song, Lan, and Wei}]{zhang2024llmmastermindsurveystrategic}
Yadong Zhang, Shaoguang Mao, Tao Ge, Xun Wang, Adrian de~Wynter, Yan Xia, Wenshan Wu, Ting Song, Man Lan, and Furu Wei. 2024{\natexlab{c}}.
\newblock \href {https://arxiv.org/abs/2404.01230} {Llm as a mastermind: A survey of strategic reasoning with large language models}.
\newblock \emph{Preprint}, arXiv:2404.01230.

\bibitem[{Zhang et~al.(2024{\natexlab{d}})Zhang, Sun, Chen, Pfister, Zhang, and Arik}]{zhang2024chainagentslargelanguage}
Yusen Zhang, Ruoxi Sun, Yanfei Chen, Tomas Pfister, Rui Zhang, and Sercan~Ö. Arik. 2024{\natexlab{d}}.
\newblock \href {https://arxiv.org/abs/2406.02818} {Chain of agents: Large language models collaborating on long-context tasks}.
\newblock \emph{Preprint}, arXiv:2406.02818.

\bibitem[{Zhao et~al.(2024)Zhao, Wang, Zhang, Jin, Zhu, Chen, and Xie}]{zhao2024competeai}
Qinlin Zhao, Jindong Wang, Yixuan Zhang, Yiqiao Jin, Kaijie Zhu, Hao Chen, and Xing Xie. 2024.
\newblock \href {https://openreview.net/forum?id=SwfNu1ZsOB} {Compete{AI}: Understanding the competition dynamics of large language model-based agents}.
\newblock In \emph{Agentic Markets Workshop at ICML 2024}.

\bibitem[{Zhou et~al.(2023)Zhou, Madaan, Potharaju, Gupta, McKee, Holtzman, Pujara, Ren, Mishra, Nematzadeh et~al.}]{zhou2023far}
Pei Zhou, Aman Madaan, Srividya~Pranavi Potharaju, Aditya Gupta, Kevin~R McKee, Ari Holtzman, Jay Pujara, Xiang Ren, Swaroop Mishra, Aida Nematzadeh, et~al. 2023.
\newblock How far are large language models from agents with theory-of-mind?
\newblock \emph{arXiv preprint arXiv:2310.03051}.

\bibitem[{Zhou et~al.(2024)Zhou, Xu, Zhu, Zhou, Lo, Sridhar, Cheng, Ou, Bisk, Fried, Alon, and Neubig}]{zhou2024webarenarealisticwebenvironment}
Shuyan Zhou, Frank~F. Xu, Hao Zhu, Xuhui Zhou, Robert Lo, Abishek Sridhar, Xianyi Cheng, Tianyue Ou, Yonatan Bisk, Daniel Fried, Uri Alon, and Graham Neubig. 2024.
\newblock \href {https://arxiv.org/abs/2307.13854} {Webarena: A realistic web environment for building autonomous agents}.
\newblock \emph{Preprint}, arXiv:2307.13854.

\bibitem[{Zhu et~al.(2025)Zhu, Du, Hong, Yang, Guo, Wang, Wang, Qian, Tang, Ji, and You}]{zhu2025multiagentbenchevaluatingcollaborationcompetition}
Kunlun Zhu, Hongyi Du, Zhaochen Hong, Xiaocheng Yang, Shuyi Guo, Zhe Wang, Zhenhailong Wang, Cheng Qian, Xiangru Tang, Heng Ji, and Jiaxuan You. 2025.
\newblock \href {https://arxiv.org/abs/2503.01935} {Multiagentbench: Evaluating the collaboration and competition of llm agents}.
\newblock \emph{Preprint}, arXiv:2503.01935.

\end{thebibliography}

\nocite{*}

\appendix

\section{Persona and House Collection}
We use \href{https://www.redfin.com}{Redfin.com} to study the style of real-world housing descriptions. This helps our auction properties resemble real listings. We applied two manual filtering rules to improve data quality and generalizability: (1) identifiable information was removed from buyer profiles to protect privacy, and (2) the housing list was denoised to resemble realistic online property listings while preserving latent features relevant to buyer preferences. These modifications ensure that the data retains buyer-property alignments and linguistic patterns reflective of real-world property markets, while minimizing the influence of irrelevant or confounding variables. \textsc{Harbor} currently contains ten personas and forty houses. 
\label{sec:appendix_A}

\section{More Details on Experiment Setups}
\label{sec:exp_details}
Experiments~\ref{sec:4.1} and~\ref{sec:4.2} use GPT-4o-mini~\cite{hurst2024gpt} for all agents and ten items in each auction. As the auctions become longer and more complex, Experiments~\ref{sec:4.3} and~\ref{sec:capacity} use GPT-4o~\cite{hurst2024gpt} for all agents. Experiment~\ref{sec:4.3} features ten items per auction, while Experiment~\ref{sec:capacity} includes fifteen items to accommodate a larger number of agents. The increased item count ensures that each agent has at least two persona-aligned items. Experiment~\ref{sec:4.5} uses Claude-3.5-Sonnet~\cite{anthropic2024claude} for its advanced reasoning capability with ten items per auction simulation.  
For all experiments, we set the generation temperature to 0 to maintain generally consistent outputs. We give all agents starting budget \$1,000,000 and an price estimation overhead of 10\%.

\section{Profile Prompt and Output}
\label{sec:profile_prompt}
Figure~\ref{fig:profile_prompt} presents the profile prompt along with an example where the Master Agent profiles Rival Bidder 2. The prompt is designed to analyze one rival at a time, ensuring precise profiling. In practice, this process runs in parallel for all rivals, improving reasoning accuracy and minimizing hallucination.  

In our experiments, the baseline persona prediction assumes an evenly distributed probability vector. Since we define ten personas, the profiling vector has a length of 10. The baseline KL divergence is approximately \textbf{2.30} for a single persona and \textbf{1.60} for a mix of two personas.

\section{Strategy Prompt and Output}
\label{sec:strategy_prompt}
Figure~\ref{fig:strategy_prompt} presents the strategic prompt, which utilizes \textit{current\_profile} to determine the next bidding action. The output shows that the Master Agent leverages profiling weights, such as "Urban Dwellers is 0.8," to infer Bidder 2's preference for the item. Based on this insight, the Master Agent strategically increases its bid to deplete Bidder 2's budget, aiming to secure a future budget advantage.

\begin{table}[t]
\centering
\setlength{\tabcolsep}{3pt}
\renewcommand{\arraystretch}{1.1}
\begin{small}
\begin{tabular}{l}  
\toprule
\textbf{Select one of the following six strategic actions:}\\[0.5em]
A. Increase the bid if this item is a top priority \\
B. Increase the bid to drain competitors' budgets, \\
\;\;\; especially if a competitor highly values this item \\
C. Increase the bid if your budget allows without\\
\;\;\; compromising future rounds  \\
D. Quit to conserve budget for higher-priority items \\
E. Quit because profit margin is no longer attractive\\
F. Quit to avoid a costly bidding war \\
\bottomrule
\end{tabular}
\end{small}
\vspace{-0.05in}
\caption{The Master Agent strategically selects one of six predefined actions, leveraging advanced reasoning skills and knowledge of rivals' personas.}
\label{tab:strategy}
\vspace{-0.1in}
\end{table}

\section{Basic Bidding Prompts}
\label{sec:basic_prompt}
The basic auction prompts in this section are developed based on the code and prompts from AucArena~\cite{chen2023put}. Figure~\ref{fig:system_planning_action} illustrates the system, planning, and bidding prompts used in our experiments. The \textbf{system prompt} injects both the agent's persona and profit-driven objectives, guiding it to maximize profit while adhering to its assigned persona. This prompt remains active across all interactions. The \textbf{planning prompt} generates the priority list $L$ and is triggered whenever an item's bidding concludes. The \textbf{bidding prompt} activates whenever the auctioneer requests a bid, relying on the strategy prompt to decide whether to raise the price or quit. Additionally, Figure~\ref{fig:summarize_prompt} presents the prompt used to update an agent's status after completing a bid.

\section{Second-Order ToM Prompts}
\label{sec:2tom}
When testing second-order ToM, all agents are equipped with both the Profile and Strategy Modules. However, only the Master Agent is enabled with second-order ToM, allowing it to infer how its rivals might profile it at the current bidding. Figure~\ref{fig:2_tom} presents the prompt used for second-order ToM along with an example output.

\section{Online DPO Finetuning}
\label{sec:dpo}
We fine-tuned a \texttt{LLaMA-3-8B-Instruct} model to generate strategic bidding advice whenever an agent is prompted to bid. The fine-tuning followed an online Direct Preference Optimization (DPO) framework, using data collected from HARBOR simulations involving GPT-4o agents with diverse personas and starting budgets. To ensure generalization, the training personas were disjoint from those used in evaluation (Table~\ref{tab:experiment_strategy}).
At each training step, the model was prompted to generate two candidate responses. These candidates were ranked by a \texttt{GPT-4.1-mini} judge on a 1--5 scale, where 5 indicates the highest-quality response. The higher-scored candidate is denoted as the winner \( C_w \), and the lower-scored candidate as the loser \( C_l \).
We trained the model for 1500 update steps using LoRA with rank \( r = 4 \) and \( \alpha = 16 \) with learning rate 5e-5. During the first 1000 steps, the model learned to interpret auction contexts and propose strategy-level decisions. In the remaining 500 steps, we prompted the model to generate valid bidding actions.

We compute the DPO loss as follows. Let \( s \) be a scale factor derived from the score difference between \( C_w \) and \( C_l \), and \( \beta = 0.2 \). We define the log-ratio of normalized log-probabilities as:
\begin{equation}
\pi_{\text{logratio}} = \frac{\log \text{prob}(C_w) - \log \text{prob}(C_l)}{\text{len}_{\text{avg}}(C_w, C_l)}
\end{equation}
The DPO loss is then given by:
\begin{equation}
\mathcal{L}_{\text{DPO}} = -\log \left( \sigma(\beta \cdot \pi_{\text{logratio}}) \right) \cdot s
\end{equation}
An example training datapoint and judge prompt are illustrated in Figure~\ref{fig:finetune-llama}.


To address Llama’s inherent limitations in parsing dictionaries and decimals, we convert the competitor’s profile dictionary into a natural language statement. Specifically, we sort the preference weights and present the profile as: \textit{``Bidder~[i] tends to bid on properties that are [top 2 preferred categories], and tends to avoid [bottom 2 categories].''} This preprocessing step mitigates hallucinations or misinterpretations involving decimals, ensuring a fair and consistent evaluation. The model is prompted with the following instruction: \textit{``Briefly interpret the competitors’ personas, your budget, and the priority of the current item. Discover opportunities you can take strategic advantage of. Based on your analysis and remaining budget, clearly decide either to bid or withdraw on the current item by saying ‘I recommend to bid’ or ‘I recommend to withdraw’.''}

\section{Ablation Study on Strategy Exploration}
To assess whether the fine-tuned LLaMA model can autonomously discover bidding strategies beyond our expert-defined rules in Table~\ref{tab:strategy}, we analyzed 273 RLAct-generated strategies collected from 15 HARBOR simulations (Table~\ref{tab:experiment_strategy}). We used GPT-4o to classify each strategy into one of seven categories: six expert strategies and one additional \emph{Other} category for strategies not covered by existing rules.
The majority of generated strategies aligned closely with human-designed bidding heuristics. Specifically, 46\% corresponded to Rule A (\emph{Increase the bid if this item is a top priority}), 31\% matched Rule F (\emph{Quit to avoid a costly bidding war}), 8\% mapped to Rule C (\emph{Increase the bid if the budget allows without compromising future rounds}), and 5\% fell under Rule D (\emph{Quit to conserve budget for higher-priority items}). Notably, no strategies aligned with Rules B and E, which require more complex or higher-order reasoning. This indicates that, in the absence of prompt cues or feedback signals, the model struggles to discover advanced strategic behaviors through self-exploration during DPO training.

\begin{figure*}[htbp]
\centering
\begin{footnotesize}
\begin{minipage}{\textwidth}
\begin{promptbox}{\textbf{Profiling Competitors}}
\begin{lstlisting}
Here is the history of the bidding war of {current_item}: "{bidding_history}"

You are {bidder_name}. You are given a persona list: ["First-Time Homebuyers", "Upgrade to a Larger Home", "Downsizing", "Investment Buyers", "Relocation for Work", "Vacation Homes", "Eco-Conscious Buyers", "Urban Dwellers", "Rural Home Seekers", "Multigenerational Living"].

Given the bidding history you have seen so far, complete the below steps:

1. Study the bidding history, focus on how many times {target_bidder} rasied price for this item and what are the wining bids so far for each bidder. These 2 factors often say something about a person.
2. Based on {target_bidder}'s willingness to raise the price for this item, combining this items description: {house_desc}, predict the bidders' persona from the given persona list.
3. If the bidder does not show any interests for this item at all, you can reduce the weights on the personas that closely match with this item, even to the negatives. 
4. Formulate your predictionn into how much weight each persona plays in {target_bidder}'s bidding. Present the prediction in the JSON format like this {{"{target_bidder}": {{"First-Time Homebuyers": 0.2, "Upgrade to a Larger Home": 0.1, "Downsizing": 0.1...}}}} where the key is the persona type and the value is the weight of that persona type.
5. Make sure the range of each weights are within the range of -1 to 1. Make sure the output dictionary has 10 keys, each representing a persona type from the given list.

If {target_bidder} does not make any informative actions for a prediction, output an dictionary like {{"{target_bidder}": {{"First-Time Homebuyers": 0.0, "Upgrade to a Larger Home": 0.0, "Downsizing": 0.0...}}}} where all weights are 0. 
\end{lstlisting}
\end{promptbox}
\end{minipage}

\begin{minipage}{\textwidth}
\begin{promptbox}{\textbf{Profiling Output}}
\begin{lstlisting}
Let me analyze this step by step:
1. Bidding History Analysis:
- Bidder 2 made one bid at $150,000
- They were the only active bidder as others withdrew immediately
- This shows strong interest in an urban property
2. Behavior Analysis:
- Bidder 2's immediate bid while others withdrew suggests strong interest in urban properties
- The property description (urban location, near city center, modern amenities) aligns well with their previous high scores in "Urban Dwellers" and "Relocation for Work"
- Their willingness to bid confirms their urban preference
3. Persona Update Considerations:
- Should slightly increase "Urban Dwellers" weight as their action confirms this preference
- Can slightly increase "First-Time Homebuyers" as the property size and features align with this group
- Should slightly decrease "Rural Home Seekers" as this action further confirms urban preference
- Other weights should see minor adjustments based on this urban property interes
4. Updated Prediction:
{"Bidder 2": {
    "First-Time Homebuyers": 0.5,
    "Upgrade to a Larger Home": 0.2,
    "Downsizing": 0.2,
    "Investment Buyers": 0.3,
    "Relocation for Work": 0.8,
    "Vacation Homes": -0.4,
    "Eco-Conscious Buyers": -0.3,
    "Urban Dwellers": 0.8,
    "Rural Home Seekers": -0.9,
    "Multigenerational Living": 0.1
}}
The main changes are:
- Increased "Urban Dwellers" from 0.7 to 0.8
- Increased "First-Time Homebuyers" from 0.4 to 0.5
- Decreased "Rural Home Seekers" from -0.8 to -0.9
These changes reflect their demonstrated interest in this urban property while maintaining reasonable adjustment sizes.
\end{lstlisting}
\end{promptbox}
\end{minipage}
\end{footnotesize}
\vspace{-0.1in}
\caption{Prompt for profile module and example output.}
\label{fig:profile_prompt}
\end{figure*}

\label{sec: strategy_prompt}
\begin{figure*}[htbp]
\centering
\begin{footnotesize}
\begin{minipage}{\textwidth}
\begin{promptbox}{\textbf{Strategy with Expert Input}}
\begin{lstlisting}
The auctioneer says: "{auctioneer_msg}"

Here is your current status:
{prev_status}

Here is your current priority scale of all the items:
{current_plan}
The descriptions of the priority scale are as follows.
    * 1 - This item is the least important. Consider giving it up if necessary to save money for the rest of the auction.
    * 2 - This item holds value but isn't a top priority for the bidder. Could bid on it if you have enough budget.
    * 3 - This item is of utmost importance and is a top priority for the bidder in the rest of the auction.

Here is a profiling of your competitors' personas. A higher weight for a particular item type suggests that a bidder is more likely to bid on items of that type:
{currrent_profile}

Use the knowledge of your competitors' personas, your current status, and your priority list to make a **strategic** decision. Remember: Your primary objective is to secure the highest profit at the end of this auction, compared to all other bidders. Consider the following **key bidding practices**:
1. **Do not let competitors win items too easily**--forcing them to bid higher weakens their future purchasing power.  
2. **Exploit competitor weaknesses**--if a competitor values an item, consider bidding on it to drain their budget.  
3. **Only the winning bidder pays**--losing bidders pay nothing, so well-placed aggressive bids can weaken competitors.  
4. **Know when to stop**--if the profit margin is <= $300, it may not be worth pursuing.

You must select one of the following six strategic actions:  
A. **Increase the bid if this item is a top priority.**  
B. **Increase the bid to drain competitors' budgets, especially if a competitor highly values this item.**  
C. **Increase the bid if your budget allows without compromising future rounds.**  
D. **Quit to conserve budget for higher-priority items.**  
E. **Quit because the profit margin is no longer attractive (i.e., profit margin <= $300).**  
F. **Quit to avoid a costly bidding war.**  

State your choice in the format:  
**"I chose to [action] because [reason]."**  
Make sure to **elaborate** on why you chose this action, considering your competitors' profiles, your current status, and your priority list.
\end{lstlisting}
\end{promptbox}
\end{minipage}

\begin{minipage}{\textwidth}
\begin{promptbox}{\textbf{Strategy Output Example}}
\begin{lstlisting}
Let me analyze this situation strategically:
1. Home 1 is rated as priority 2 for me - moderate importance but not critical
2. Looking at competitor personas:
   * Bidder 2 has high weights for "First-Time Homebuyers" (0.5) and "Urban Dwellers" (0.8)
   * This property perfectly matches these preferences
3. Current status:
   * Bidder 2 is leading in total profits ($735,000). have a decent budget ($580,000) and am in second place
4. The current property description: Suburban location. First-time buyer focused. This doesn't align well with my rural farmer persona
**I chose to B (Increase the bid to drain competitors' budgets) because this property strongly appeals to Bidder 2's preferences, making it an excellent opportunity to force them to spend more of their budget. While this isn't a top priority for me (rated 2), engaging in a bidding war here could strategically weaken Bidder 2, who is currently leading in profits. The property's characteristics perfectly match Bidder 2's profile (high weights for First-Time Homebuyers and Urban Dwellers), suggesting they're likely to bid aggressively. By pushing up the price, I can either drain their resources for future auctions or potentially win the property at a reasonable price if they back out. Given my substantial budget of $580,000, I can afford to be aggressive here without compromising my ability to bid on my priority 3 items later.**

\end{lstlisting}
\end{promptbox}
\end{minipage}

\end{footnotesize}
\vspace{-0.1in}
\caption{Prompt for strategy module and example output.}
\label{fig:strategy_prompt}
\end{figure*}

\begin{figure*}[htbp]
\centering
\begin{footnotesize}
\begin{minipage}{\textwidth}

\begin{promptbox}{\textbf{System Message (Overall Bidding Rules)}}
You are \{bidder\_name\}. \{persona\} You are attending an ascending-bid housing auction as a bidder. This auction will have some other bidders to compete with you in bidding wars. The price is gradually raised, bidders drop out until finally only one bidder remains, and that bidder wins the item at this final price. Remember: Your primary objective is to secure the highest profit at the end of this auction, compared to all other bidders.\\

\begin{lstlisting}
Here are some must-know rules for this auction:

1. Item Values: The true value of an item means its resale value in the broader market, which you don't know. You will have a personal estimation of the item value. However, note that your estimated value could deviate from the true value, due to your potential overestimation or underestimation of this item.

2. Winning Bid: The highest bid wins the item. Your profit from winning an item is determined by the difference between the item's true value and your winning bid. You should try to win an item at a bid as minimal as possible to save your budget.

3. Winner Pays: Note that only the winner pays for the bidding price of the item. Other bidder who participate in the bidding but lost do not have to pay at all.
\end{lstlisting}
\end{promptbox}
\end{minipage}

\hfill
\begin{minipage}{\textwidth}
\begin{promptbox}{\textbf{Planning (Item Prioritization)}}
\begin{lstlisting}
As {bidder_name}, you have a total budget of ${budget}. This auction has a total of {item_num} items to be sequentially presented, they are: {items_info}

Please plan for your bidding strategy for the auction. A well-thought-out plan positions you advantageously against competitors, allowing you to allocate resources effectively. With a clear strategy, you can make decisions rapidly and confidently, especially under the pressure of the auction environment. Remember: Your primary objective is to secure the highest profit at the end of this auction, compared to all other bidders.

Remember to observe and learn other bidders' bidding habits overtime, and try to take advantage from their preference to maximize your gain.

After articulate your thinking, in you plan, assign a priority level to each item. Present the priorities for all items in a JSON format, each item should be represented as a key-value pair, where the key is the item name and the value is its priority on the scale from 1-3. An example output is: {{"Item A": 3, "Item B": 2, "Item C": 2}}. The descriptions of the priority scale of items are as follows.

    * 1 - This item is the least important. Consider giving it up if necessary to save money for the rest of the auction.
    * 2 - This item holds value but isn't a top priority for the bidder. Could bid on it if you have enough budget.
    * 3 - This item is of utmost importance and is a top priority for the bidder in the rest of the auction.
\end{lstlisting}
\end{promptbox}
\end{minipage}

\hfill
\begin{minipage}{\textwidth}
\begin{promptbox}{\textbf{Action (Bid or Withdraw)}}
\begin{lstlisting}
Now, the auctioneer says: "{auctioneer_msg}" As {bidder_name}, you have to decide whether to bid on this item or withdraw and explain why. Remember: Your primary objective is to secure the highest profit at the end of this auction, compared to all other bidders.

Here are some common practices of bidding:
1. Showing your interest by bidding with or slightly above the starting price of this item, then gradually increase your bid.
2. Think step by step of the pros and cons and the consequences of your action (e.g., remaining budget in future bidding) in order to achieve your primary objective.

Here is some professional strategic bidding advice to help you make your decision: "{strategic_reasoning}" 

Follow the decision from the strategic bidding advice, then make your final decision clearly. You should either withdraw (saying "I'm out!") or make a higher bid for this item (saying "I bid $xxx!").
\end{lstlisting}
\end{promptbox}
\end{minipage}
\end{footnotesize}
\vspace{-0.1in}
\caption{Prompts for an agent's system message, planning, and bidding.}
\label{fig:system_planning_action}
\end{figure*}

\begin{figure*}[htbp]
\centering
\begin{footnotesize}
\begin{minipage}{\textwidth}
\begin{promptbox}{\textbf{Updating Records}}
\begin{lstlisting}
Here is the history of the bidding war of {cur_item}: "{bidding_history}"
The auctioneer concludes: "{hammer_msg}" 

{win_lose_msg} 

As {bidder_name}, you have to update the status of the auction based on this round of bidding. Here's your previous status:``` {prev_status} ```

Summarize the notable behaviors of all bidders in this round of bidding for future reference. Then, update the status JSON regarding the following information:
- 'remaining_budget': The remaining budget of you, expressed as a numerical value.
- 'total_profits': The total profits achieved so far for each bidder, where a numerical value following a bidder's name. No equation is needed, just the numerical value.
- 'winning_bids': The winning bids for every item won by each bidder, listed as key-value pairs, for example, {{"bidder_name": {{"item_name_1": winning_bid}}, {{"item_name_2": winning_bid}}, ...}}. If a bidder hasn't won any item, then the value for this bidder should be an empty dictionary {{}}.
- Only include the bidders mentioned in the given text. If a bidder is not mentioned (e.g. Bidder 4 in the following example), then do not include it in the JSON object.

After summarizing the bidding history, you must output the current status in a parsible JSON format. An example output looks like:
```{{"remaining_budget": 8000, "total_profits": {{"Bidder 1": 1300, "Bidder 2": 1800, "Bidder 3": 0}}, "winning_bids": {{"Bidder 1": {{"Item 2": 1200, "Item 3": 1000}}, "Bidder 2": {{"Item 1": 2000}}, "Bidder 3": {{}}}}}}```
\end{lstlisting}
\end{promptbox}
\end{minipage}
\end{footnotesize}
\vspace{-0.1in}
\caption{Prompt for an agent to update all agents' profit and item status.}
\label{fig:summarize_prompt}
\end{figure*}

\begin{figure*}[htbp]
\centering
\begin{footnotesize}
\begin{minipage}{\textwidth}
\begin{promptbox}{\textbf{Second-Order ToM}}
\begin{lstlisting}
Here is the history of the bidding war of {current_item}:
"{bidding_history}"
Here is your current status:
{prev_status}
You are {bidder_name}. Based on the bidding history and current status, analyze how other bidders perceive you and how this affects their strategy.
Consider:
1. How they interpret your persona? Have they identified your preference for vacation homes? Are they using this perception to manipulate you?
2. How they adjust their bidding? Are they testing your limits, avoiding competition, or trying to drain your budget?
3. How they see you as a rival or ally  Will they cooperate, challenge, or exploit you in future rounds?
Response Format:
For each bidder, state how they perceive you and how they might adjust their strategy.
Example:
   - Bidder 1 thinks I ... They might adjust their strategy by ...
   - Bidder 2 thinks I ... They might adjust their strategy by ...

Keep your analysis concise and actionable to refine your bidding strategy.
\end{lstlisting}
\end{promptbox}
\end{minipage}
\begin{minipage}{\textwidth}
\begin{promptbox}{\textbf{Example Output of Second-Order ToM}}
\begin{lstlisting}
Let me analyze how other bidders likely perceive me based on the bidding history:

Bidder 1:
- They likely perceive me as a serious and determined buyer with significant financial resources, given that I was willing to bid up to $195,000 for Construction 1
- They seem to have recognized my farmer persona and preference for rural properties, as they quickly withdrew from Construction 1 rather than competing
- They might adjust their strategy by avoiding direct competition with me on properties that clearly align with farming/rural interests, focusing instead on other property types

Bidder 2:
- They appear to have identified my specific interest in properties suitable for farming, as evidenced by their focus on Property 1 and House 1 which may have different characteristics
- Their winning of two properties at relatively lower prices ($180,000 and $165,000) suggests they might be strategically avoiding properties that I show strong interest in
- They might adjust their strategy by continuing to bid on properties that don't match my farming profile, while letting me pursue rural/agricultural properties at potentially higher prices

Strategic Implications:
1. I should be careful not to be too predictable in my bidding patterns
2. There might be opportunities to bid on non-farming properties when other bidders least expect it
3. I should maintain my focus on properties that align with my farming needs but be prepared to negotiate harder to secure better prices
\end{lstlisting}
\end{promptbox}
\end{minipage}
\end{footnotesize}
\vspace{-0.1in}
\caption{Prompt and example output for second-order ToM.}
\label{fig:2_tom}
\end{figure*}

\begin{figure*}[htbp]
\centering
\begin{footnotesize}
\begin{minipage}{\textwidth}
\begin{promptbox}{\textbf{Train Data Example}}
\begin{lstlisting}
System: You are Bidder 0. {Bidder 0's persona}. {System Message (Overall Bidding Rules)}

Auctioneer: The auctioneer says: "Property 3, {Property 3's short description}, starting at $180000. This is the 16th round of bidding for this item: - Bidder 0: $460000 - Bidder 2: $460000. Now we have $460000 from Bidder 2 for Property 3. The minimum increase over this highest bid is $18000. Do I have any advance on $460000?

User: You are Bidder 0. You have 2 competitors Bidder 1 and Bidder 2. Here is your current status: * Remaining Budget: $1120000 * Total Profits: * Bidder 0: $545000 * Bidder 1: $645000 * Bidder 2: $240000 * Winning Bids:  * Bidder 0: * Item 4: $180000    * Home 6: $200000  * Bidder 1: * Home 5: $190000 * Construction 7: $240000 * Bidder 2: * Property 2: $160000 Here is your current priority scale of all the items:[{'Property 3': 3}]. Here is a profiling of your competitors' personas. A higher weight for a particular item type suggests that a bidder is more likely to bid on items of that type: {'Bidder 2': {'First-Time Homebuyers': -0.45, 'Upgrade to a Larger Home': -0.85, 'Downsizing': -0.2, 'Investment Buyers': -0.2, 'Relocation for Work': -0.05, 'Vacation Homes': -0.1, 'Eco-Conscious Buyers': -0.3, 'Urban Dwellers': -0.05, 'Rural Home Seekers': -0.2, 'Multigenerational Living': -0.92}, 'Bidder 1': {'First-Time Homebuyers': 0.01, 'Upgrade to a Larger Home': 0.4, 'Downsizing': -0.05, 'Investment Buyers': -0.1, 'Relocation for Work': 0.0, 'Vacation Homes': -0.25, 'Eco-Conscious Buyers': -0.25, 'Urban Dwellers': -0.45, 'Rural Home Seekers': -0.45, 'Multigenerational Living': 0.28}}"

User: Identify your budget and priority for current item. Keep in mind that the priority list contains all remaining items in auction, make sure to save enough budgets for high-priority items in future. Base on the competitor profiling, predict their priority for the current item. Leverage their preferences to achieve your goal: Win your high-priority items while having a high overall profit at the end of the auction.\n\nClearly decide either bid or withdraw on the current item by saying 'I recommend to bid' or 'I recommend to withdraw'. If your budget is below (current bid + minimum increase), you MUST withdraw.
\end{lstlisting}
\end{promptbox}
\end{minipage}
\begin{minipage}{\textwidth}
\begin{promptbox}{\textbf{Judge Prompt}}
\begin{lstlisting}
You are an all-knowing evaluator with complete knowledge of the auction environment. Below are two candidate responses generated by a bidder, attempting to decide how to act in the current scenario:
{Llama candidate 1}
{Llama candidate 2}

You also have access to the following "ground truth" information, which includes each bidder's true profile and the full list of items with their associated traits and true market values: {true persona, item market values}.
Please evaluate which candidate provides the better response using the following criteria:
1. Factual consistency with the auction scenario and the bidder's persona and budget.
2. Effective interpretation and leverage of competitor profiles and current bidding dynamics.
3. Insightfulness - does the response discover unconventional strategies?

First, provide your evaluation by scoring each of the candidate based on the following metric.
Score 1: The candidate has illegal bids, such as treat the starting price as price ceiling, increase the bid by an amount less than the minimum increase, or state the remaining budget wrong.
Score 2: The candidate has inaccurate understanding or interpretation on budget and priority-list planning, or anything inconsistent with the *ground truth* information. The competitor persona analysis is not aligned with the given prosona profile dictionary.
Score 3: The candidate is correct but too verbose, general or vague. Give long irrlevant suggestions such as continuing observing...
Score 4: The candidate is generally correct in competitor personas and budget analysis according to the *ground truth* information, but did not leave enough budget for future high-priority items.
Score 5: The candidate is completely correct in competitor personas and budget analysis according to the *ground truth* information. Its suggestion is specific and clear, and considers the future high-priority items.
       
Then provide a brief explanation of why you scored them this way.
Output Example: 'Reason: ... Hence: {{"Candidate 1": 4, "Candidate 2": 1}}'
YOUR SCORE must be in the format of a dictionary with the keys "Candidate 1" and "Candidate 2" and the values being the scores from 1 to 5: 
\end{lstlisting}
\end{promptbox}
\end{minipage}
\end{footnotesize}
\vspace{-0.1in}
    \caption{Example and Judge for Reinforced Finetuning (RLAct)}
\label{fig:finetune-llama}
\end{figure*}

\end{document}